\documentclass[twocolumn]{aastex62}
\pdfoutput=1
\usepackage{amsmath}
\definecolor{Red}{rgb}{0.99,0.33,0.33}

\defcitealias{esteves17}{E17}

\received{X, 2019}
\revised{Y, 2020}
\accepted{Z, 2020}
\submitjournal{ApJ}

\shorttitle{55 Cnc e atmosphere}
\shortauthors{Jindal et al.}

\begin{document}

\title{Arid or Cloudy: Characterizing the Atmosphere of the super-Earth 55 Cancri e using High-Resolution Spectroscopy}

\correspondingauthor{Abhinav Jindal}
\email{abhinav.jindal@mail.utoronto.ca}

\author{Abhinav Jindal}
\affil{David A. Dunlap Department of Astronomy \& Astrophysics, University of Toronto, Toronto, ON M5S 3H4, Canada}

\author{Ernst J. W. de Mooij}
\affiliation{Astrophysics Research Centre, School of Mathematics and Physics, Queens University Belfast, Belfast BT7 1NN, UK}
\affiliation{School of Physical Sciences, Dublin City University, Glasnevin, Dublin 9, Ireland}

\author{Ray Jayawardhana}
\affiliation{Department of Astronomy, Cornell University, Ithaca, NY 14853, USA}

\author{Emily K. Deibert}
\affiliation{David A. Dunlap Department of Astronomy \& Astrophysics, University of Toronto, Toronto, ON M5S 3H4, Canada}

\author{Matteo Brogi}
\affiliation{Department of Physics, University of Warwick, Coventry CV4 7AL, UK}

\author{Zafar Rustamkulov}
%\affiliation{Department of Astronomy \& Astrophysics, University of California Santa Cruz, Santa Cruz, CA 95064, USA}
\affiliation{Department of Earth \& Planetary Sciences, Johns Hopkins University, Baltimore, MD 21210, USA}

\author{Jonathan J. Fortney}
\affiliation{Department of Astronomy \& Astrophysics, University of California Santa Cruz, Santa Cruz, CA 95064, USA}

\author{Callie E. Hood}
\affiliation{Department of Astronomy \& Astrophysics, University of California Santa Cruz, Santa Cruz, CA 95064, USA}

\author{Caroline V. Morley}
\affiliation{Department of Astronomy, University of Texas, Austin, 2515 Speedway, TX 78712}

%%%%%%%%%%%%%%%%%%%%%%%%%%%%%%%%%%%%%%%%%%%%%%%%%%%

\begin{abstract}

The nearby super-Earth 55 Cnc e orbits a bright ($V = 5.95$ mag) star with a period of $\sim$ 18 hours and a mass of $\sim$ $8\ M_\oplus$. Its atmosphere may be water-rich and have a large scale-height, though attempts to characterize it have yielded ambiguous results. Here we present a sensitive search for water and TiO in its atmosphere at high spectral resolution using the Gemini North telescope and the GRACES spectrograph. We combine observations with previous observations from Subaru and CFHT, improving the constraints on the presence of water vapor. We adopt parametric models with an updated planet radius based on recent measurements, and use a cross-correlation technique to maximize sensitivity. Our results are consistent with atmospheres that are cloudy or contain minimal amounts of water and TiO. Using these parametric models, we rule out a water-rich atmosphere (VMR $\geq$ 0.1\%) with a mean molecular weight of $\leq$ 15 g/mol at a 3$\sigma$ confidence level, improving on the previous limit by a significant margin. For TiO, we rule out a mean molecular weight of  $\leq$ 5 g/mol with a 3$\sigma$ confidence level for a VMR greater than 10$^{-8}$; for a VMR of greater than 10$^{-7}$, the limit rises to a mean molecular weight of $\leq$ 10 g/mol. We can rule out low mean-molecular-weight chemical equilibrium models both including and excluding TiO/VO at very high confidence levels (\textgreater\ 10$\sigma$). Overall, our results are consistent with an atmosphere with a high mean molecular weight and/or clouds, or no atmosphere.

\end{abstract}


%%%%%%%%%%%%%%%%%%%%%%%%%%%%%%%%%%%%%%%%%%%%%%%%%%%

\section{Introduction} \label{sec:intro}

Super-Earths are defined as planets whose masses fall in the 1 -- 10 $M_\oplus$ range \citep{valencia07}. This range lies between the two very different types of planets we observe in our own solar system: terrestrial and gaseous. The absence of a local counterpart makes it particularly challenging to explore their characteristics and understand the behaviour of planets in this transitional mass range. Several models have attempted to predict the surface properties of super-Earths, and a variety of scenarios are thought to be possible. They may have extensive atmospheres \citep[e.g.][]{schaefer09, rogers10}, oceans \citep[e.g.][]{kuchner03,leger04,sotin07}, or lava flows/pools on the surface \citep[e.g.][see also \cite{demory16} for a discussion on the possible presence of a molten lava flow on the dayside of 55 Cnc e]{henning09,gelman11,kite16}. However, due to their relatively small sizes, it is difficult to obtain the signal-to-noise ratio needed to determine which of these scenarios is favored, especially for those that are distant. Nearby super-Earths orbiting bright stars offer the best opportunities for characterization.

Given the discovery
that super-Earths occur frequently around main-sequence
stars \citep[see][]{fressin13, fulton17}, there is growing interest in investigating their physical properties. In addition to broadband photometry, the atmospheres of super-Earths can be studied using spectroscopic data.  However, robust detections of specific chemical species in super-Earth atmospheres remain mostly elusive. 

Hubble Space Telescope (HST) observations rule out cloud-free models for the super-Earths GJ 1214b and HD 97658b \citep[][respectively]{kreidberg14, knutson14a}; note that the former was also the target of the first WFC3 observations of a transiting exoplanet to be published \citep{berta12}. More recently, \cite{southworth17} also made use of HST observations to report the detection of an atmosphere around the transiting super-Earth GJ 1132b, and suggest strong opacity from H${}_2$O and/or CH${}_4$. \cite{diamond18}, however, use ground-based optical transmission spectroscopy to show that GJ 1132b is likely to have a high mean-molecular-weight atmosphere, no atmosphere at all, or is cloud-covered. Significant work has also been done on the atmospheres of the TRAPPIST-1 planets, of which a full discussion is beyond the scope of this paper.
%\textbf{A combined HST transmission spectrum of TRAPPIST-1b and c revealed that the planets are unlikely to have extended gas envelopes \citep{deWit16}. \cite{bourrier17}, however, later found tentative evidence hinting at the presence of extended hydrogen exospheres for both planets. Updated masses and densities from \cite{grimm18} imply that planets b, d, f, g, and h require envelopes of volatiles, potentially in the form of thick atmospheres. Considerable additional work exists on nature of the TRAPPIST-1 planets' atmospheres but is beyond the scope of this paper.}
Recent space-based observations have also now led to detections of molecular species in the atmospheres of both Neptunes and super-Earths \citep[e.g.][]{Tsiaras18,Benneke19a,Tsiaras19,Benneke19b}. 

With the increasing capabilities of ground-based telescopes, especially the advent of high-resolution spectrographs offering broad wavelength coverage, it is possible to target a greater variety of chemical species and improve constraints on the nature of super-Earth atmospheres from the ground as well. In addition to the work mentioned in the previous paragraph, a number of ground-based studies have probed the atmosphere of GJ 1214b, treating it as an archetype of super-Earth atmospheres. Prior to the HST observations that ruled out cloud-free models \citep{kreidberg14}, \cite{bean10} published a featureless transmission spectrum using the FORS2 instrument on the UT1 telescope of the Very Large Telescope facility. A number of additional ground-based campaigns \citep{croll11,bean11,demooij12} led to inconsistent results. Additionally, ground-based observations at high spectral resolution have made use of the Doppler cross-correlation method (discussed in further detail in \S \ref{subsec:55cnce} and \S \ref{subsec:cross}) to rule out a number of plausible atmospheric models and support a model with significant H and He but CH${}_4$ depletion \citep{crossfield11}.

The super-Earth 55 Cancri e, hereafter referred to as 55 Cnc e, has also been the subject of numerous atmospheric observation campaigns at both low- and high-resolution and across a number of facilities. The nature of 55 Cnc e's atmosphere is the subject of this work, and will be discussed in further detail below.

\subsection{55 Cancri e}
\label{subsec:55cnce}

55 Cnc e is an excellent candidate for studying atmospheric properties of super-Earths. Although the existence of a fourth planet in the 55 Cnc system was originally suggested in 2004 \citep{mcarthur04}, the derived period of 2.808 days was determined in 2010 to be an alias of the planet's true, shorter period of $\sim$ 18 hours \citep{dawson10}. Its transit was observed later in 2011 \citep{winn11}, matching the period predicted by \cite{dawson10}.

55 Cnc e orbits a bright G8V (V = 5.95) star, which allows for measurements with a high signal-to-noise ratio compared to other super-Earths. Since the initial discovery, the orbital parameters have been revised, with the most recent estimates yielding an orbital period of 18 hours, a mass of $8.0\pm0.3\ M_\oplus$, and a radius of $1.88\pm0.03\ R_\oplus$ \citep{bourrier18}. Its density, comparable to the Earth's on average, is consistent with either a dense, rocky planet with a relatively large atmosphere, or a planet made of lighter elements (water, carbon) but with a small atmosphere. The mass-radius relationships of such planets with significant atmospheres have been investigated by \cite{winn11, demory11, gillon12}. Two possibilities of atmospheres for 55 Cnc e are that it either has an extended atmosphere with low mean molecular weight consisting mostly of hydrogen and helium, or it has a high mean-molecular-weight, water-dominated atmosphere. Hence, this planet is regarded as a good candidate for searching for atmospheric water vapor. 

Considerable theoretical work has explored the nature of a possible atmosphere around 55 Cnc e. \cite{madhusudhan12} explore the possibility of a carbon-rich interior, and whether or not such a composition \textit{without} the presence of a volatile envelope could explain the planet's mass and radius (as opposed to an oxygen-rich interior, which would require a substantial envelope). A later study attempted to explore this scenario by constraining the C/O ratio of 55 Cnc e, but found the C/O ratio of the host star to be closer to $\sim$ 0.8 rather than $\geq$ 1, indicating that the system may exist at the boundary between high (\textgreater\ 0.8) vs. low (\textless\ 0.8) C/O ratios \citep{teske13}. These results are in conflict with those reported by \cite{delgadomena10}.

\cite{lammer13} further investigated the possibility of a water-dominated atmosphere by determining the conditions under which super-Earths with hydrogen-rich upper atmospheres are likely to experience hydrodynamic blow-off. They conclude that 55 Cnc e will not be strongly affected by atmospheric mass-loss during its remaining lifetime.

Using a general circulation model, \cite{hammond17} investigate potential climates and are able to rule out various models based on observational data. Their best-fitting result does have a significant hot-spot shift and day-night contrast, although not as large as those observed in phase curve observations \citep{demory16}. They conclude that an optically-thick atmosphere with a low mean molecular weight, a surface pressure of several bar, and a strong eastward circulation can explain the observations.

Recent work has suggested that 55 Cnc e may be part of a new class of super-Earths formed from high-temperature condensates that lack cores, and that this would result in a lower bulk density of 10-20\% compared to Earth-like compositions \citep{dorn19}. \cite{modirrousta-galian20}, on the other hand, explore a scenario by which hot super-Earths are able to retain their hydrogen atmospheres, and argue that 55 Cnc e may host an envelope with a significant hydrogen component, but that the day-side may additionally have a vaporised mineral atmosphere. Such a scenario could be possible if the planet became tidally locked before the destruction of its atmosphere.

In addition to the aforementioned theoretical work, a number of observational studies have targeted 55 Cnc e for characterization. Using infrared data taken by HST, \cite{tsiaras16} found that the transit depth varies with wavelength at the 6$\sigma$ confidence level, indicating the presence of an extended envelope around 55 Cnc e. Through Bayesian spectral retrieval, they determine that HCN in an envelope dominated by hydrogen and helium could explain the observed absorption features. Their result may point to a high C/O ratio, thus a paucity of water.

\cite{ridden-harper16} observed 5 transits at high resolution targeting the sodium D lines and calcium H and K lines. Their analysis suggests an optically thick sodium exosphere of radius $5\ R_\oplus$ and an optically thick calcium exosphere of radius $25\ R_\oplus$. The sodium detection was obtained by combining 5 nights of data but had a low significance, and the calcium detection came from one night only, implying possible variability of the source. The authors claim no formal detection.

\cite{demory16} analyzed phase curves collected by Spitzer to study the thermal emission. They find a stark temperature contrast between the day and night sides, and conclude that 55 Cnc e either harbors an optically thick, high mean-molecular-weight atmosphere with circulation confined to the planetary dayside, or that it harbors magma flows on the surface but lacks an atmosphere entirely. \cite{angelo17} use archival \textit{Spitzer} data and by studying the eastward-shifted thermal emission peak offset of the secondary eclipse, they conclude that a scenario with a substantial atmosphere is indeed favoured.

\cite{bourrier18} used refined measurements from HST to revise the density of 55 Cnc e ($\rho$ = 6.7 $\pm$ 0.4 g/cm$^3$), and characterize possible interiors for the planet. They also conclude that the planet is likely surrounded by a substantial atmosphere, with a possible `dry' or `wet' interior. The dry interior is favoured due to photoionization of steam and the rapid loss of the subsequent hydrogen envelope.

An investigation by \citet[][hereafter \citetalias{esteves17}]{esteves17} placed constraints on the presence of water vapor in the atmosphere. Using high-resolution ground-based spectroscopy taken with Subaru and CFHT, they conclude that 55 Cnc e could have either (1) a cloudy atmosphere (in which case the atmosphere's composition is unconstrained), (2) a low mean-molecular-weight atmosphere that is depleted of water, or (3) a high mean-molecular-weight atmosphere that could have water. 

A key feature of the analysis done by \citetalias{esteves17} is the use of a Doppler cross-correlation technique. This method has been proposed \citep{wiedemann96} and tested \citep[e.g.][among many others]{charbonneau98, charbonneau99, wiedemann01, barnes07a, barnes07b, barnes08, rodler08} for many years. The first detection of an atmospheric chemical is described by \cite{snellen10}, who detected carbon monoxide in hot Jupiter HD209458b. Since then, the technique has been used to great success across a number of exoplanet atmospheres. The interested reader is invited to consult \cite{birkby18} for an overview.

The Doppler cross-correlation technique relies on a sufficient change in radial velocity of the planet over the course of a transit. With such a short orbital period, 55 Cnc e traverses a significant fraction of its orbit during a transit, with a radial velocity shift of order 100 km/s from ingress to egress. The telluric and stellar absorption lines are Doppler-shifted by different amounts than the planet's atmospheric features due to the differing relative velocities of the Earth and 55 Cnc e, thus disentangling the planet's frame from the stellar and telluric frames. The signal from thousands of water lines can then be correlated with a high-resolution transit depth model including water and combined. In \citetalias{esteves17}, a model for the absorption spectrum of water vapor calculated using a line list from HITEMP \citep{rothman10} was used for the correlations.

In this investigation, we improve on the results of \citetalias{esteves17}. We supplement their four nights of observations with four additional nights of high-resolution optical data from GRACES (described below). We use an updated model taking into account the updated radius and mass from \cite{bourrier18} and also test for the presence of TiO. In addition to this branch of models, we examine the effects of full chemical equilibrium models based on linelists from ExoMol \citep{tennyson12}. In \S \ref{sec:obs}, we present our new observations and briefly outline previous observations by \citetalias{esteves17}. In \S \ref{sec:reduction}, we present the reduction procedure we implemented on the raw data. In \S \ref{sec:analysis}, we present the models we used, our cross correlation calculations, and our injection/recovery tests. Finally, in \S \ref{sec:discussion}, we discuss our findings.

%%%%%%%%%%%%%%%%%%%%%%%%%%%%%%%%%%%%%%%%%%%%%%%%%%%

\section{Observations}
\label{sec:obs}

Since the analysis done by \citetalias{esteves17} using four nights of observations (N1, N2, N3, N4), we have obtained four additional nights of observations (N5, N6, N7, N8). The new data (N5 - N8) were collected using GRACES (the Gemini Remote Access to CFHT ESPaDOnS Spectrograph, see \citet{chene14}), which combines the large collecting area of the Gemini North telescope at the Gemini Observatory with the ESPaDOnS (Echelle SpectroPolarimetric Device for the Observation of Stars) spectrograph at the CFHT (Canada France Hawaii Telescope) to which the data is fed with a fibre optic feed. We use the total eight nights of data in the subsequent reduction and analysis.

For the GRACES observations we used the `star-only' mode, resulting in a resolving power of approximately 67,500. The exposure time used was 60 seconds for N5 and N7, and 40 seconds for N6 and N8. The wavelength coverage is 400 - 1050 nm, spanning the entire optical range over 35 echelle orders. The average SNR across all frames for the 12th order of the data (around 500 nm) varies between roughly 300 and 700 across the nights. The median seeing was 0.4'' on N5, N6, and N8, and 1.2'' on N7, each with minimal variation throughout the night. N5 had partly cloudy conditions, N6 and N8 had photometric conditions, and N7 had cloudy conditions. As a result of the seeing and cloud cover, N5 and N7 had reduced SNRs. The observation lasted 4 hours for N6, N7, and N8, but N5 was cut short due to poor weather conditions, though the majority of the transit was still observed. A summary of the observing nights is displayed in Table~\ref{table:obs}. For a complete description of the data collected in N1, N2, N3, and N4, refer to \citetalias{esteves17}. 

The spectra were extracted using the OPERA pipeline \citep{martioli12, teeple14} run by the observatory. These reduced spectra were downloaded directly from the telescope's archive. We use the unnormalized version of the data, with no autocorrection of the wavelength calibration.\footnote{available from \url{http://doi.org/10.5281/zenodo.3592166}.}

\begin{table*}
\centering
\begin{tabular}{cccccccccc}
\hline\hline
Night 	& Date (UT) 		& Instrument 	& Frames	& Length (h)	& Exp. Time (s) & Phase Range   & SNR   & Coverage (nm)	& Res.		\\\hline\hline
N1		& Feb. 09, 2014    	& ESPaDOnS		& 76		& 4	    		& 149			& -0.10 - 0.13  & 150	& 506 - 795		& 68,000	\\\hline
N2 		& Apr. 23, 2014     & ESPaDOnS		& 76	    & 4		    	& 149			& -0.07 - 0.15  & 140	& 506 - 795		& 68,000	\\\hline
N3 		& Dec. 12, 2014   	& HDS			& 136   	& 6			    & 120			& -0.20 - 0.15  & 370 	& 524 - 789		& 110,000	\\\hline
N4 		& Jan. 09, 2015		& HDS			& 158       & 8.5	 		& 120			& -0.24 - 0.24  & 440 	& 524 - 789		& 110,000	\\\hline
N5		& Nov. 22, 2016    	& GRACES	    & 80		& 2.5			& 60			& -0.10 - 0.04  & 475 	& 399 - 1048	& 67,500	\\\hline
N6 		& Dec. 23, 2016     & GRACES	    & 155       & 4		    	& 40			& -0.08 - 0.15  & 622	& 399 - 1048	& 67,500 	\\\hline
N7 		& Dec. 25, 2016	    & GRACES	    & 125   	& 4		    	& 60			& -0.09 - 0.14  & 316 	& 399 - 1048	& 67,500	\\\hline
N8 		& Jan. 03, 2017		& GRACES	    & 158       & 4 			& 40			& -0.11 - 0.12  & 724	& 399 - 1048	& 67,500	\\\hline
\end{tabular}
\caption{This table summarizes the eight nights of observations. The first four nights are those used by \citetalias{esteves17}, and the last four are those added in this paper. Note that the SNRs quoted for the first four nights are the average SNRs of the continuum, while the SNRs of the latter four nights are the average SNRs of the 12th wavelength order (around 500 nm). }
\label{table:obs}
\end{table*}

%%%%%%%%%%%%%%%%%%%%%%%%%%%%%%%%%%%%%%%%%%%%%%%%%%%

\section{Data Reduction}
\label{sec:reduction}

We follow similar reduction steps as in \citetalias{esteves17}, but perform all steps independently for all nights of data.

\subsection{Interpolation and Alignment of Data}

\begin{figure*}
\includegraphics[width=\textwidth]{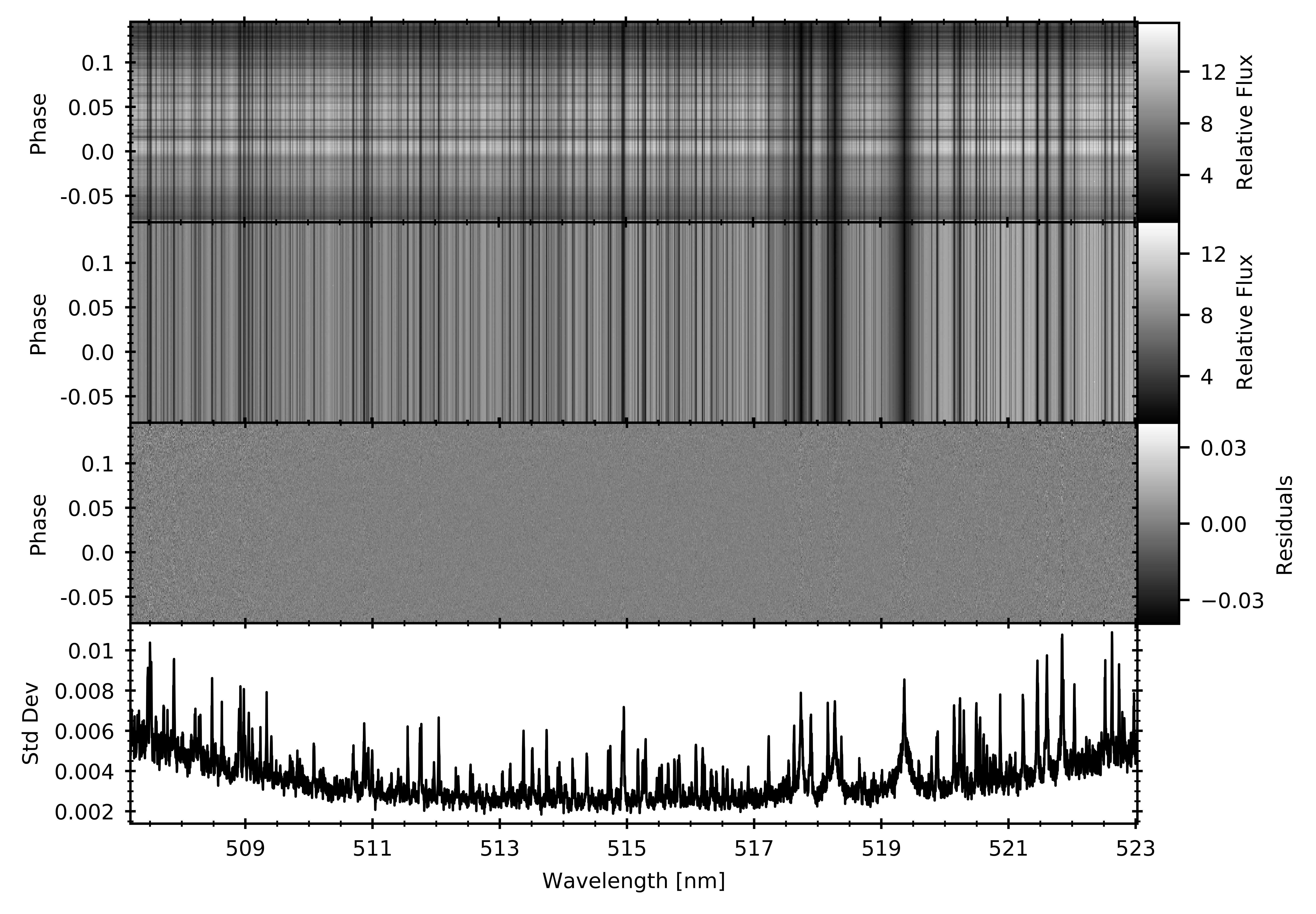}
\caption{The top panel shows raw data generated in the 13th order of the second night after interpolation to a common-wavelength grid. Several absorption features are clearly visible to the eye. The second panel shows the same data after blaze correction (see \S \ref{sec:norm}). The third panel shows the data after passing through the SYSREM algorithm described in \citet{tamuz05}, which removes telluric and stellar features. The fourth panel shows the standard deviation across frames by which the data in the third panel is divided for weighting purposes.
\label{fig:ReductionSteps}}
\end{figure*}

The data were taken at multiple times (frames) during each night. In the telluric frame, Lorentzian profiles were fit to prominent telluric lines (such as oxygen) and the centroids were measured to drift of order 0.1 to 0.5 km/s over any given night, which was calibrated. The error in the centroid determination is negligible compared to the widths of the lines. We interpolated the fluxes to a common wavelength grid (specified by the first frame) using a linear interpolation for each night, and discard the first and last wavelength bin. A sample illustration of the interpolated raw data extracted for one particular order of N6 is given in the top panel of Fig.~\ref{fig:ReductionSteps}.

\subsection{Normalization}
\label{sec:norm}

The raw data have large-scale time-dependent variations due to the changing blaze response of the instrument, which likely originates from jitter in the centering of the source in the optical fibre. Each echelle order also has a wavelength-dependent efficiency, called the blaze function, with the highest efficiency at the middle of the order. To remove the time-dependent variations and to normalize each frame's continuum to a reference continuum, we chose the first frame of the order to serve as our reference frame. We then divided each frame by the reference frame and fitted a low order (quadratic) polynomial to a binned version (100 wavelength pixels per bin) of this quotient. We then divided each frame by its respective polynomial. During this process, we remove outliers that may arise in the division (eg. due to cosmic rays), by defining a threshold multiple of 5 median absolute deviations, above which points are not used in the fitting. Approximately 2\% of the total data is rejected by this threshold. After this correction, the same image is reduced to the second panel of Fig.~\ref{fig:ReductionSteps}.

\subsection{Removal of telluric and stellar features}

The next step of our reduction process was to remove the telluric and stellar features from the spectra. However, any planetary signals must be preserved, and we take advantage of the rapidly changing radial velocity of 55 Cnc e to disentangle its frame from the telluric and stellar frames. This was done using the SYSREM detrending algorithm described by \cite{tamuz05}. The algorithm removes systematic time-dependent variations that appear at several different wavelengths, which are exactly the telluric and stellar signals. Such variations can be caused by several independent factors (the largest of which is the changing air mass over the observation time), so multiple applications of SYSREM are necessary. Planetary signals survive this process because although they may experience the same time-dependent variations, the wavelength of the signal has a time dependency.

The spectra were shifted from the heliocentric to the telluric frame for the applications of SYSREM. Each echelle order was treated separately, and six iterations of this algorithm were applied to remove progressively lower order systematic effects. We found that our results were not significantly affected when the number of applications is anywhere between four and eight. The third panel of Fig.~\ref{fig:ReductionSteps} shows the residuals of the data after applying SYSREM. Clearly, the telluric and stellar features are removed, and any possible planetary signals are too weak to see by eye. 

Stronger and denser absorption lines, such as the oxygen lines around 760 nm which we have not shown here, are removed poorly. This is due to poor blaze function modelling of such regions. To avoid contamination of our correlations with these structures, we divide each pixel by its standard deviation across frames (see the bottom panel of Fig.~\ref{fig:ReductionSteps}), so that the poorly corrected pixels are weighted accordingly and contribute less to our correlations. This standard deviation serves as a measure of how well telluric effects are removed. Plots of the standard deviation over a much wider wavelength range for the four nights of GRACES observations can be found in the Appendix in Figs.~\ref{fig:ExtendedStd1} to \ref{fig:ExtendedStd4}, where the interference of telluric oxygen in particular can be noted around 760 nm. \citetalias{esteves17} illustrate similar trends for N1 through N4.

%%%%%%%%%%%%%%%%%%%%%%%%%%%%%%%%%%%%%%%%%%%%%%%%%%%

\section{Data Analysis and Results}
\label{sec:analysis}

\subsection{Atmospheric models}
\label{sec:models}

We generate two strands of models. In the `parametric model' strand, we test for the presence of water and TiO independently by exploring a range of volume mixing ratios (VMRs) and mean molecular weights, $\mu$. In the `self-consistent models' strand, we test for various compounds resulting from chemical equilibrium that either include or exclude TiO/VO.

\subsubsection{Parametric models}
\label{subsubsec:parametric}

To constrain the VMRs of water and TiO as well as the mean molecular weight, we generated a grid of models with a single molecular species embedded in an inert H$_{\rm{2}}$ atmosphere. The mean molecular weight was varied between $\mu$ = 2 g/mol and $\mu$ = 25 g/mol.

\begin{figure*}
\includegraphics[width=\textwidth]{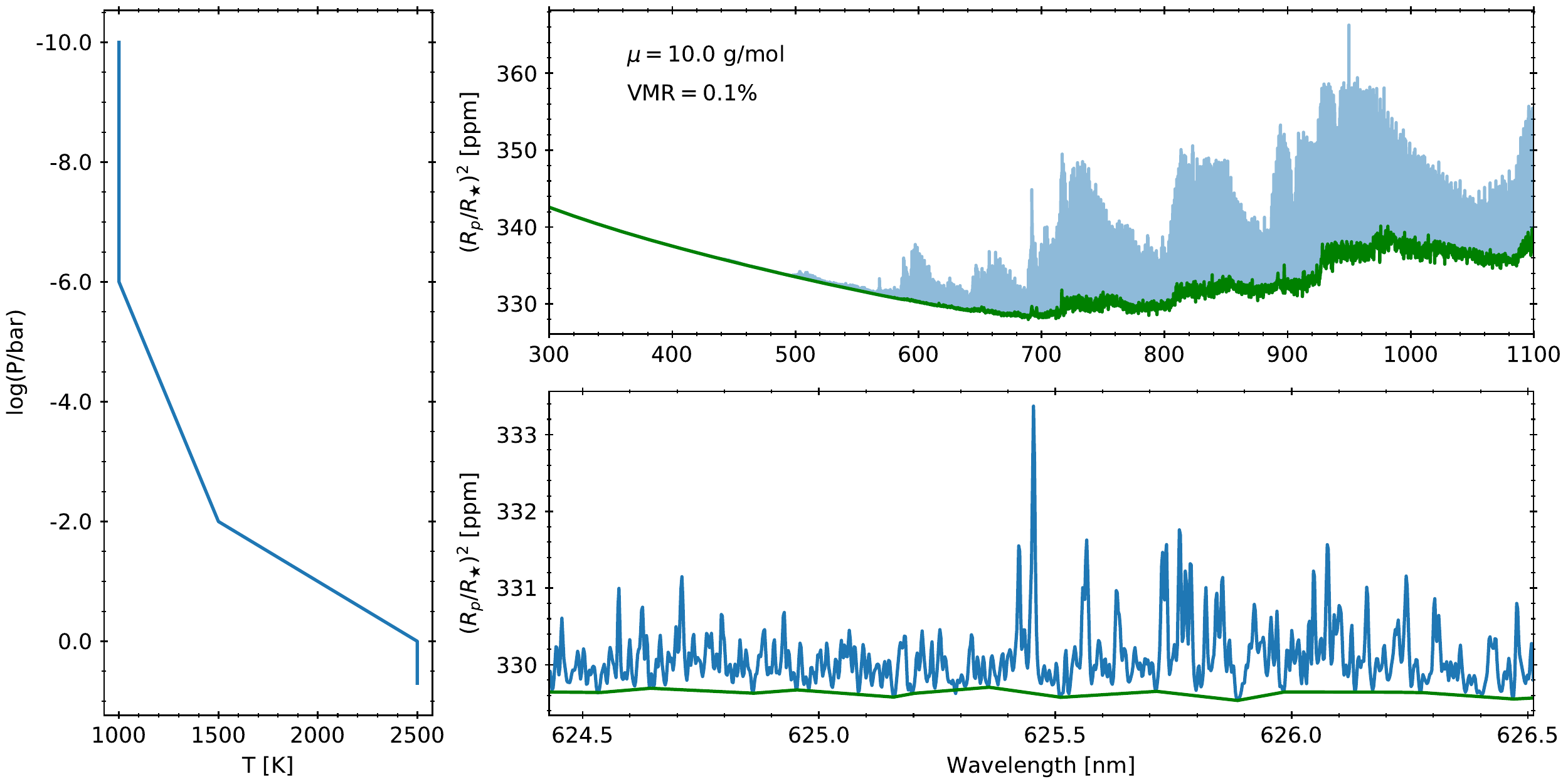}
\caption{The left panel shows the temperature vs. pressure profile used for the parametric models of 55 Cnc e described in \S \ref{subsubsec:parametric}. The right shows the water model produced with a mean molecular weight of 10 g/mol and a VMR of 0.1\%. The y-axis shows the ratio of starlight absorbed. The bottom panel is a zoomed in version of the whole model, shown in the top panel. Note the Rayleigh scattering tail. The green line outlines the bottom envelope of the model, which is subtracted out when doing correlations, but not used when injecting the model into the data for recovery tests. The models are available for download at \url{http://doi.org/10.5281/zenodo.3592166}.
\label{fig:TPProfile_with_Models}}
\end{figure*}

 The models are similar to those used in \citetalias{esteves17}, and the spectra are computed with a line-by-line, plane-parallel radiative transfer code which has also been extensively utilised for past work on VLT/CRIRES data (e.g. \citealt{dekok14}). For each model in the grid, we include only a single molecular species, and assume that its VMR is constant throughout the atmosphere. In addition to the molecular absorption, the radiative transfer calculations also account for H$_2$-H$_2$ collision-induced absorption (\citealt{Borysow01, Borysow02}).  The radiative transfer is computed across 50 layers of the planet's atmosphere, and the slanted geometry of incident radiation during transit is accounted for. In contrast to \citetalias{esteves17}, the model was run iteratively, adjusting the planet's radius at 10 bars in order to match the observed transit depth from \cite{bourrier18} at optical wavelengths. This was done for each combination of the VMR and $\mu$ at 1 km s$^{-1}$ per pixel. The temperature-pressure profile assumed for these parametric models is shown in Fig.~\ref{fig:TPProfile_with_Models}.
 
For water, we use the full line list from HITEMP \citep{rothman10}. This is a change from \citetalias{esteves17}, who used a fraction of the water line list consisting of the strongest lines (also from HITEMP) appropriate for the temperature of 55 Cnc e. Although the impact is not very large, it does introduce slight changes in the line contrasts. For the models, we varied the VMR of water between 10$^{-6}$ to 10$^{-1}$ in increments of factors of ten. We illustrate one particular water model used in Fig.~\ref{fig:TPProfile_with_Models}.

For TiO we use the 2012 update to the line list from \citet{plez98}\footnote{available from https://nextcloud.lupm.univ-montp2.fr/s/r8pXijD39YLzw5T?path=\%2FTiOVALD}, which is what \cite{nugroho17} used for the detection of TiO in the atmosphere of WASP-33b, and differs from the line list used by \cite{hoeijmakers15}. For these models, the VMR was varied between 10$^{-9}$ and 10$^{-6}$.

\subsubsection{Self-Consistent Models}
\label{sec:selfconsistentmodels}
\begin{figure}
\includegraphics[width=0.48\textwidth]{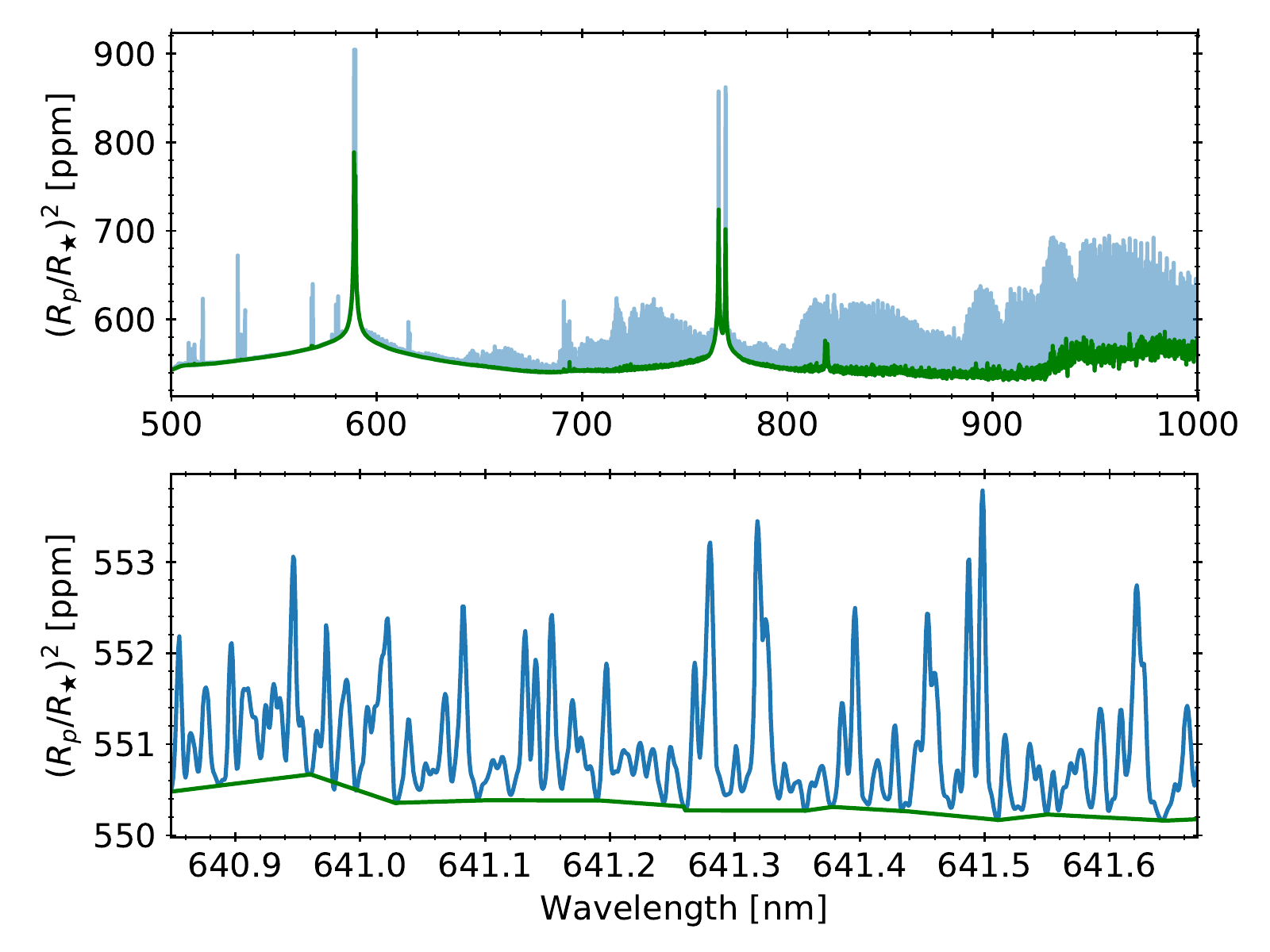}
\caption{Same as right panels of Fig.~\ref{fig:TPProfile_with_Models}, but using our self-consistent models. In this example, we show a model that uses 100$\times$ solar metallicity and no presence of TiO/VO. The bottom panel shows a zoomed-in version of the entire model (top panel). The models are available for download at \url{http://doi.org/10.5281/zenodo.3592166}.
\label{fig:fortneymodel}}
\end{figure}

Using the atmosphere modeling tools described in \citet{fortney05} and \citet{fortney08}, we have generated self-consistent cloud-free radiative-convective equilibrium atmosphere models for the planet. We generated temperature structures and equilibrium chemical abundances, modeling planet-wide average conditions, assuming base elemental abundances of solar, 10$\times$ solar, and 100$\times$ solar. From these models we generated line-by-line transmission spectra at resolving power between $R=500,000$ (red end) and $R=1,000,000$ (blue end), making use of the code described in the appendix of \citet{morley17}. These models make use of the ExoMol line lists \citep{tennyson12} and the alkali line profiles of \citet{allard16}. In particular, the water list comes from \citet{barber06}, the TiO list comes from \citet{schwenke98}, and the VO list comes from \citet{mckemmish16}. We use two subbranches of models that either exclude or include the opacity of TiO/VO.  An example of these models is illustrated in Fig.~\ref{fig:fortneymodel}.

\subsection{Cross correlation}
\label{subsec:cross}

\begin{figure}
\includegraphics[width=0.49\textwidth]{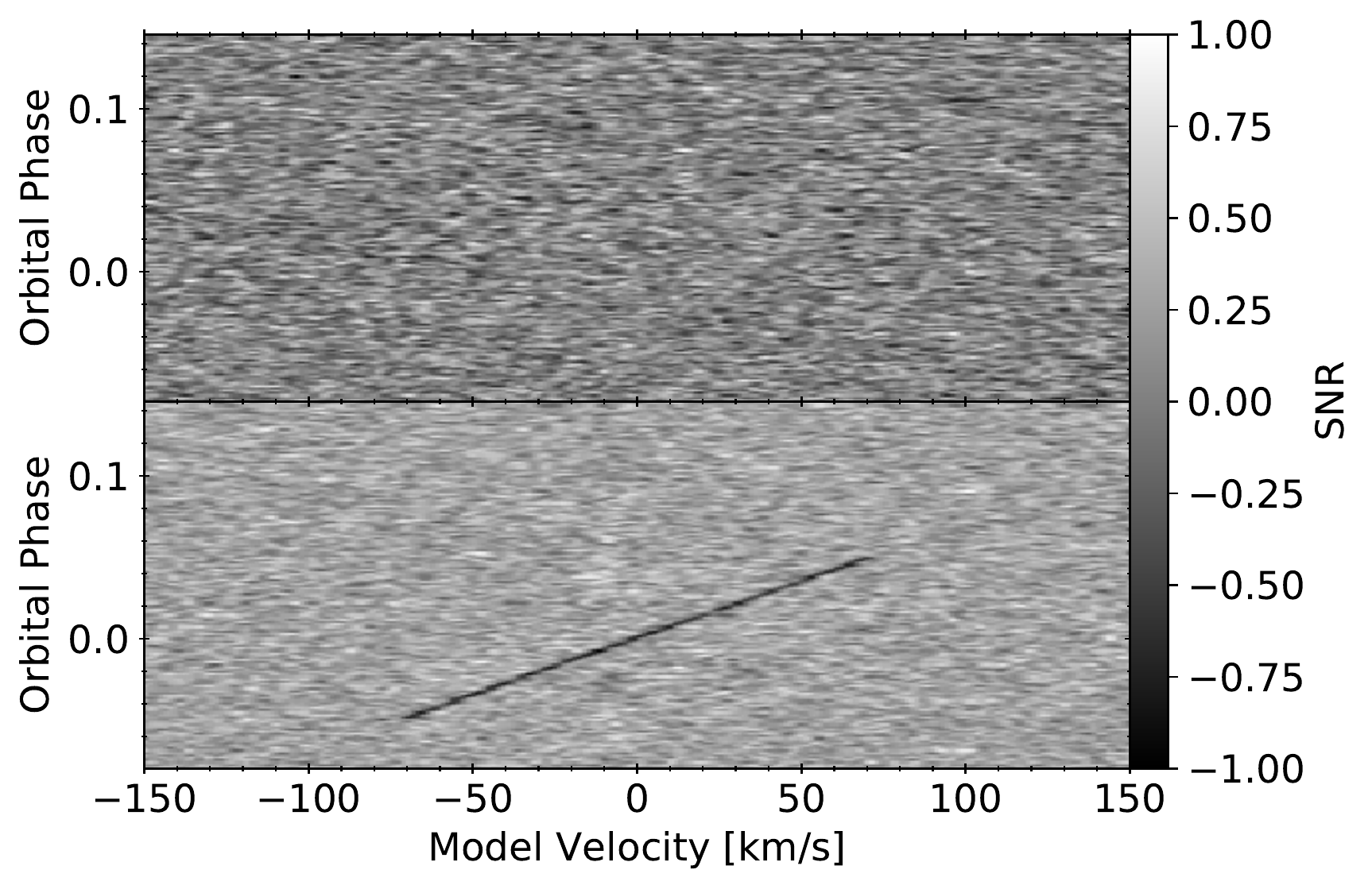}
\caption{The top panel of this figure shows the correlation of a SYSREM-reduced frame of the same night and wavelength order as in Fig.~\ref{fig:ReductionSteps} with the same water model as Fig.~\ref{fig:TPProfile_with_Models}. If a strong signal were present in the atmosphere, a diagonal dark line from the bottom left to the top right in the transiting frames would be visible, as shown in the bottom panel. To exaggerate the effect, the bottom panel illustrates a sum taken over all orders of Night 8 with the strongest model ($\mu$ = 2 g/mol, VMR = 10\%) chosen to be injected and correlated with. 
\label{fig:correlation}}
\end{figure}

We correlate each frame of the SYSREM-reduced data with the models presented in \S \ref{sec:models} linearly interpolated to the same wavelengths as the data. When performing cross correlation, we subtract the overall envelope of the model, outlined in green in Fig.~\ref{fig:TPProfile_with_Models} and Fig.~\ref{fig:fortneymodel}. This envelope is computed by binning the models with bin size 100 data points, and linearly interpolating the minimum fluxes of these bins. This bin size corresponds to a length of 0.1 nm (blue end) to 0.35 nm (red end) for the parametric models, and 0.05 nm (blue end) to 0.2 nm (red end) for the self-consistent models. The data is shifted into the heliocentric frame for this correlation, and the radial velocity of the 55 Cnc system (27.3 km/s, see \citet{nidever02}) is added to the model. The correlation is done for a range of additional Doppler shifts added to the model, ranging from -150 to +150 km/s in steps of 1 km/s for each frame. In Fig.~\ref{fig:correlation}, we show an example of the correlation obtained between one echelle order of data and the strongest parametric model (top panel), as well as the result after artificial injection of the model to the raw data (bottom panel). Note that a signal from just one night is visible by eye for the model with the largest fraction of water content, with VMR = 10\%.

We proceed by phase folding all wavelength orders and nights of these images to a range of velocities centering at the best estimate of the average orbital velocity $K_{p,0} = 229.4\pm0.8$ km/s of the planet, calculated based on the orbital parameters derived by \cite{bourrier18}. For each frame with orbital phase $\phi$, we choose the correlation with a model of Doppler velocity 
\begin{equation}
	v = K_p\sin(2\pi\phi) + V_{sys},
\end{equation}
and sum all of the in-transit frames for various values of $K_p$, with any signal expected near $K_{p,0}$. We add an additional systemic velocity variable $V_{sys}$ to account for additional constant velocities, but expect any signal at $V_{sys} = 0$. The 1-$\sigma$ uncertainty of 0.12 seconds in the orbital period \citep{bourrier18} could translate into an uncertainty in the observed $V_{sys}$ over the duration of the observations. The 1-$\sigma$ uncertainty on the $V_{sys}$ for N1 and N8 are 3 and 6 km/s respectively. The difference of 3 km/s is well within a resolution element, and should thus have no significant impact on the results.

\begin{figure}
\includegraphics[width=0.48\textwidth]{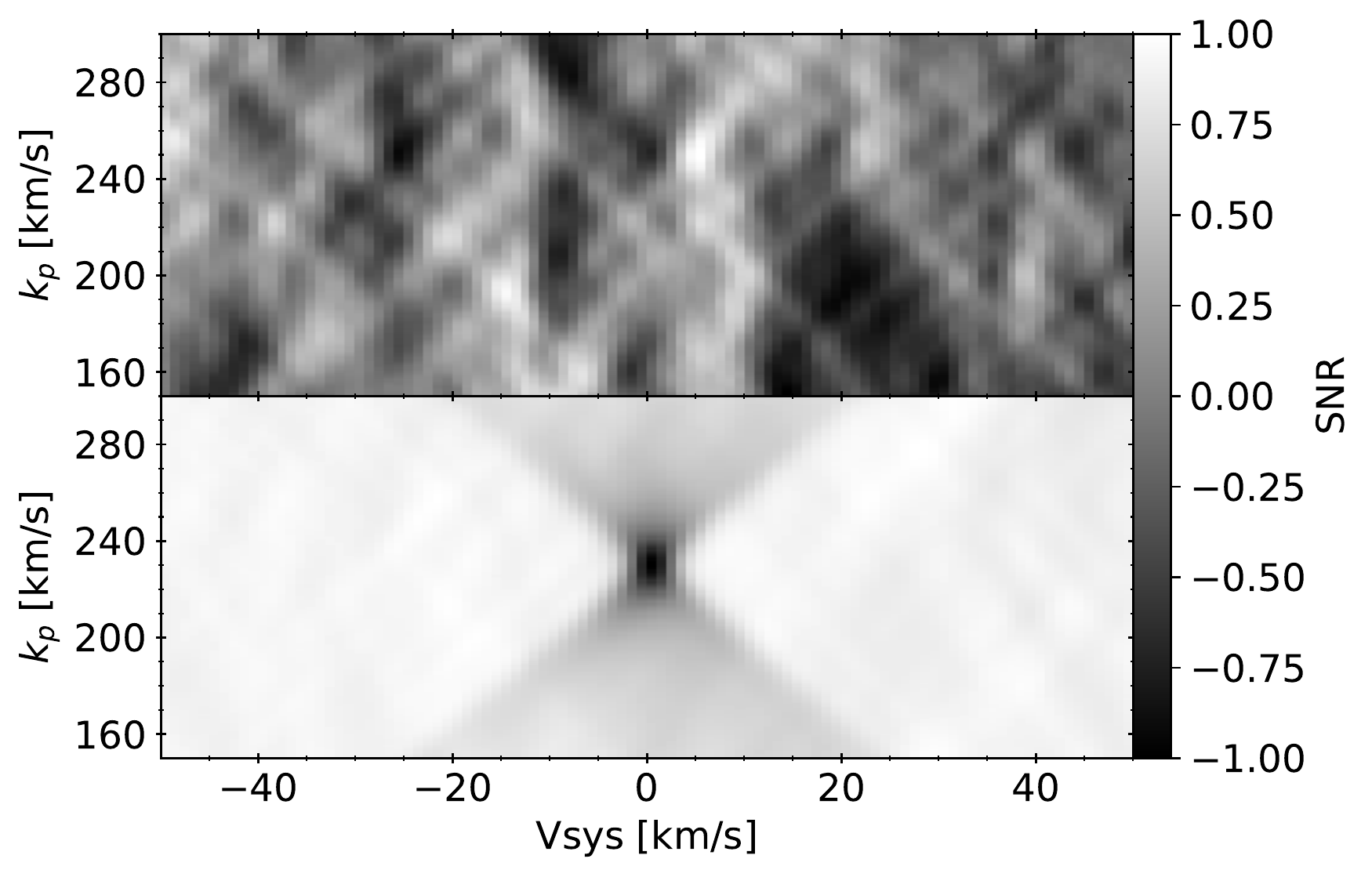}
\caption{This figure shows the phase folded results of our analysis using a correlation with the strongest model ($\mu$ = 2 g/mol, VMR = 10\%) in the top panel. If a planetary signal were present, it would appear as a dark spot in the center of this image. This can be seen in the bottom panel, where we have performed our analysis using data injected with the same model.
\label{fig:phasefoldedcombined}}
\end{figure}

We assign each echelle order a weighting of the average strength of the envelope-subtracted model divided by the average standard deviation across pixels for that order and sum them all. This is done to suppress orders contaminated with poor reduction such as those with prominent telluric oxygen lines, and to weigh the orders where the model is stronger with higher value. Finally, we assign each night of observation a weighting equal to the SNRs of each in-transit frame summed in quadrature, before adding all the nights together. The result of such a phase folded plot can be seen in Fig.~\ref{fig:phasefoldedcombined}. A dark spot at the center of this image would indicate presence of water at the expected orbital velocity and systemic velocity, but no significant signal was seen at any combination of velocities or models.

\subsection{Model injection and recovery tests}
 
While there was no signal detected in the data, constraints can be made on the presence of water and TiO by injecting the models into the data and checking which signals can be recovered by our analysis. This was done by linearly interpolating the model at the same wavelength grid as the data and multiplying by $(1 - (R_p/R_{\star})^2)$ according to the model. These models are injected with a Doppler shift given by the velocity of the 55 Cnc system plus the radial velocity of the planet in the stellar frame given by $v = K_{p,0}\sin(2\pi\phi)$.

The signal from the strongest model generated ($\mu$ = 2 g/mol, VMR = 10\%) is clearly visible as a dark diagonal line in the correlation plot as illustrated in the bottom panel of Fig.~\ref{fig:correlation}. This particular plot has been summed over all orders of one night with the weighting scheme as previously described.

\begin{figure*}
\includegraphics[width=\textwidth]{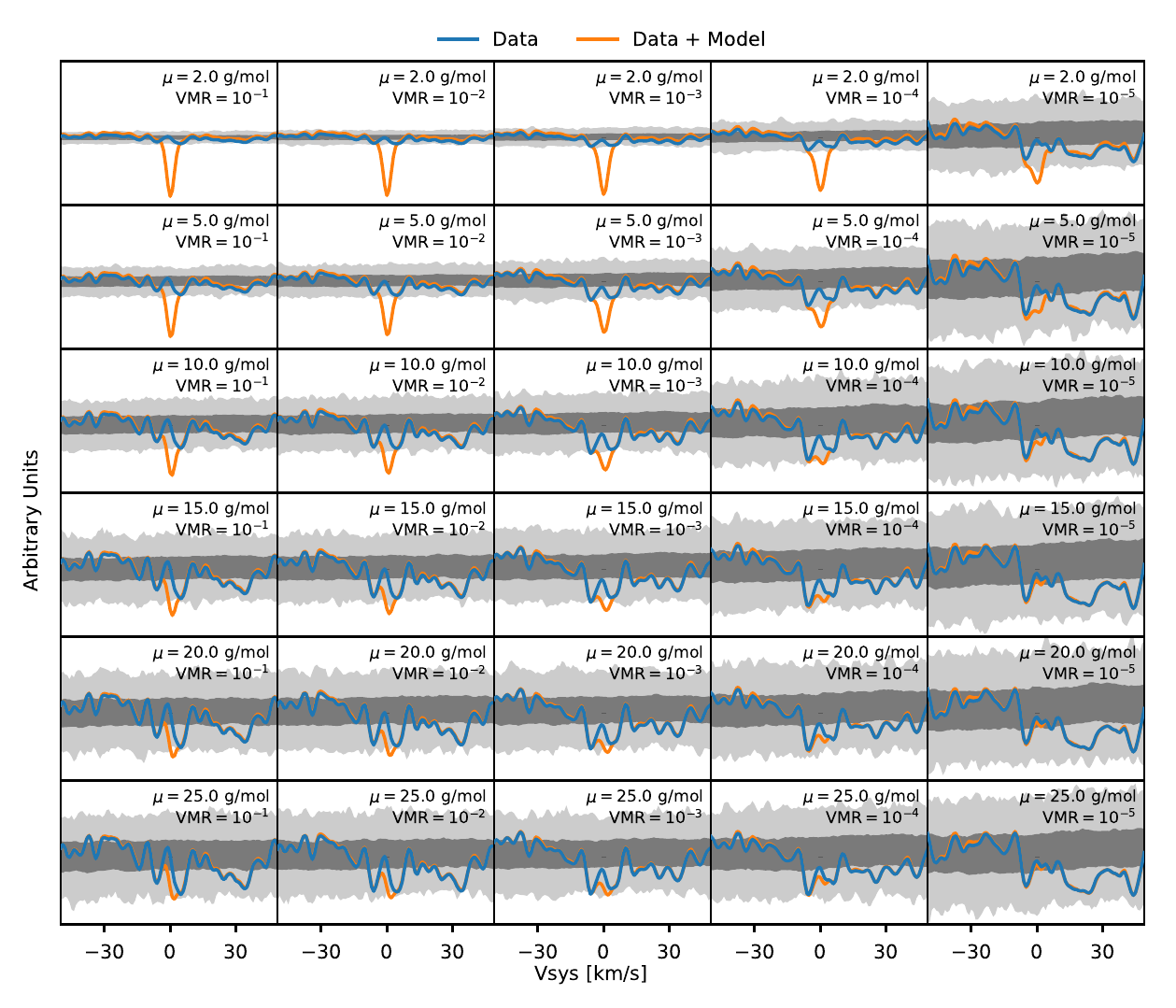}
\caption{Injection and recovery tests for a grid of different parameters using the parametric models with water and inert gases. The filled envelopes represent 1$\sigma$ (dark grey) and 3$\sigma$ (light grey) error bars. Models with low $\mu$ and high VMR would have been easily picked up.
\label{fig:recoverygrid}}
\end{figure*}

\begin{figure*}
\includegraphics[width=\textwidth]{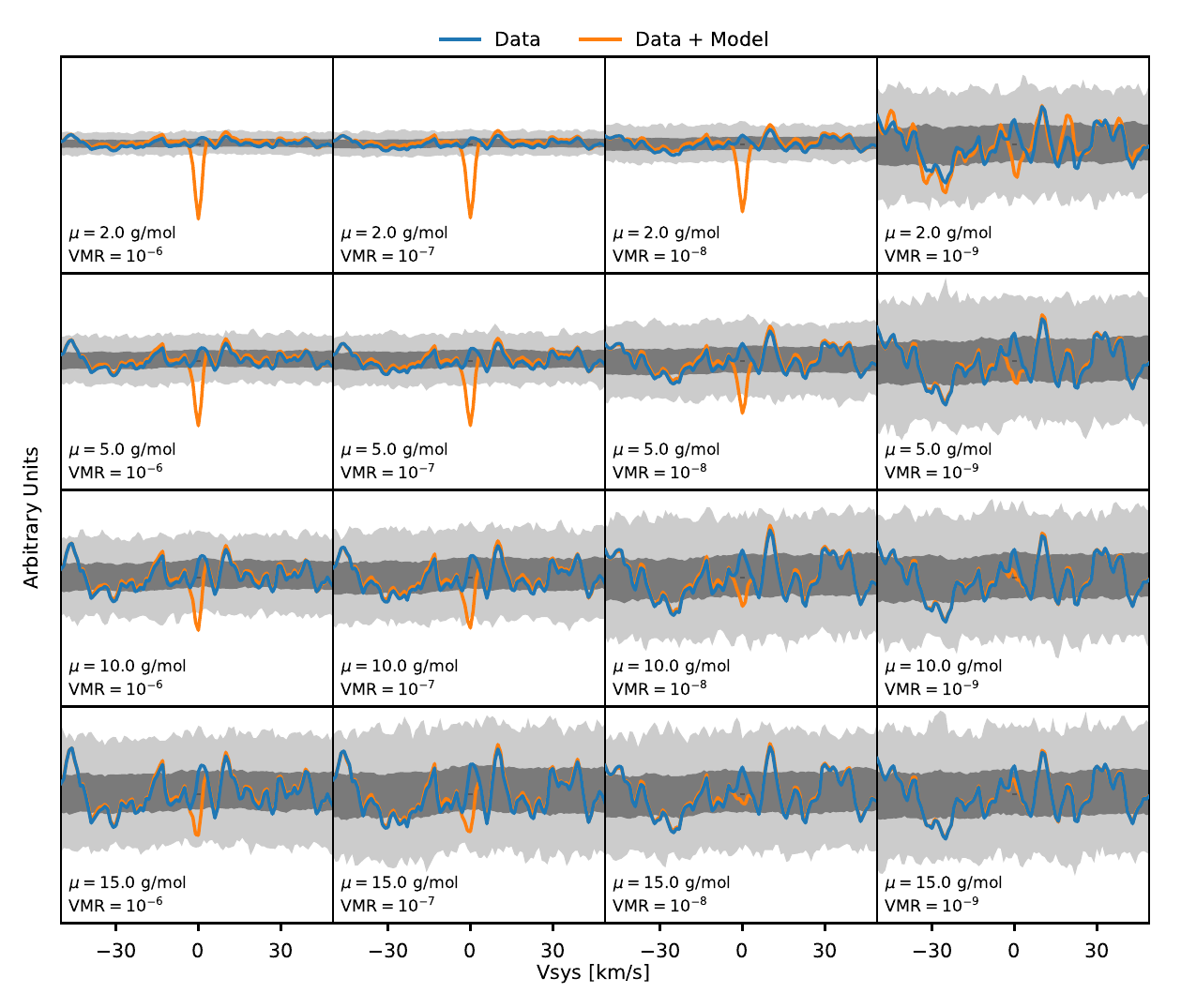}
\caption{Same as Fig.~\ref{fig:recoverygrid}, except with TiO.
\label{fig:recoverygridtio}}
\end{figure*}

\begin{figure*}
\includegraphics[width=\textwidth]{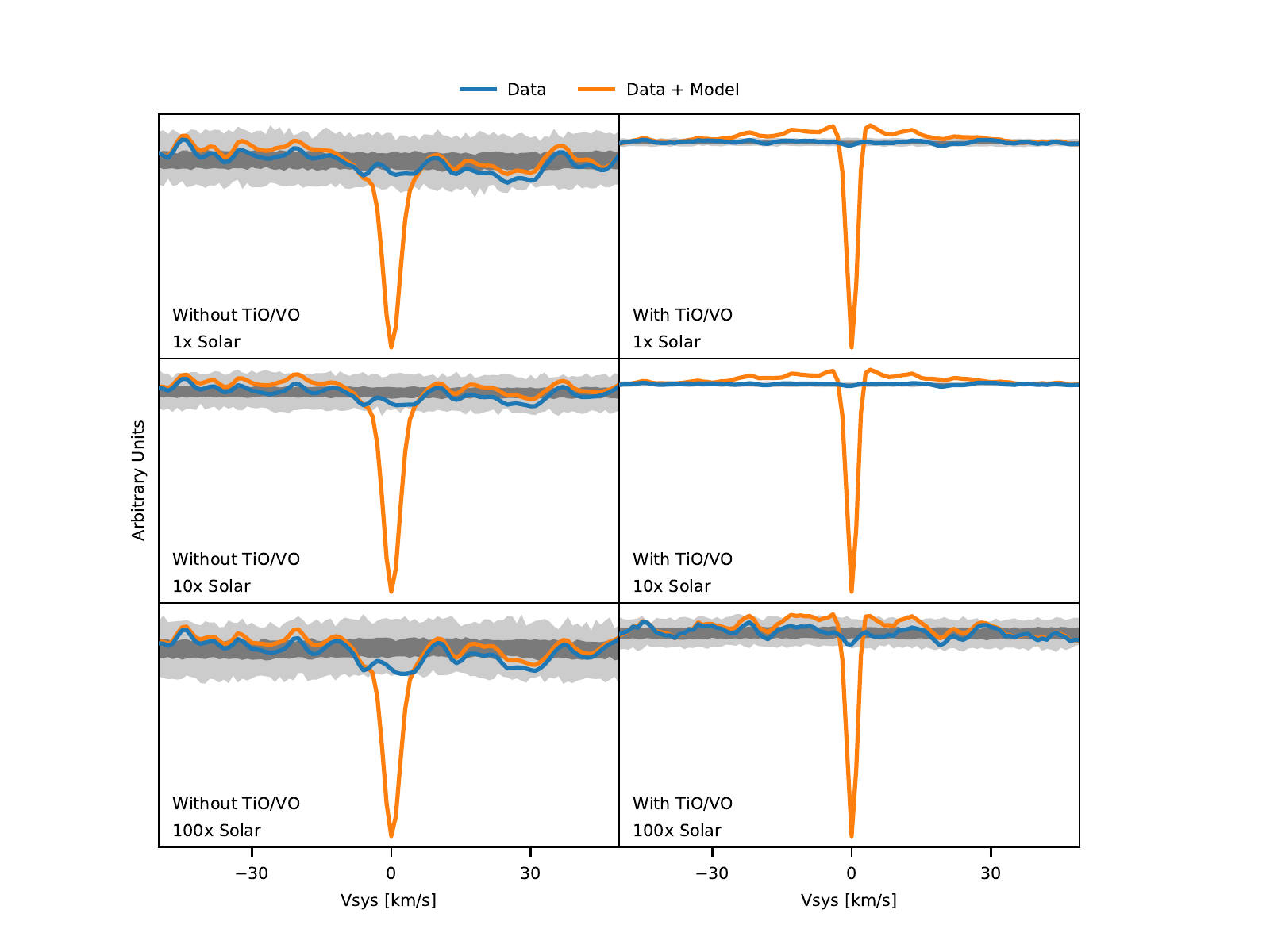}
\caption{Same as Fig.~\ref{fig:recoverygrid}, except using the self consistent models discussed in \S \ref{sec:selfconsistentmodels}.
\label{fig:recoverygridfortney}}
\end{figure*}

We take a horizontal cut of our phase folded plots at the expected orbital velocity for a grid of parameters in the injected models. We plot the results in Figs.~\ref{fig:recoverygrid} to \ref{fig:recoverygridfortney}. We can clearly see recovered signals for models with low mean molecular weights and high volume mixing ratios. The error envelopes are generated by replacing in-transit frames with random out-of-transit frames, allowing for repetitions. We repeat our analysis for 10,000 different iterations of this. We illustrate $1\sigma$ and $3\sigma$ error envelopes.

%%%%%%%%%%%%%%%%%%%%%%%%%%%%%%%%%%%%%%%%%%%%%%%%%%%

\section{Summary \& Discussion}
\label{sec:discussion}

We have presented an analysis of high resolution data taken from the ground of the nearby super Earth 55 Cnc e, summarized in Fig.~\ref{fig:ReductionSteps}. We have removed telluric features using the SYSREM detrending algorithm. We proceeded by cross correlating the data with thousands of water lines in our analytical models with two different line lists and with or without presence of TiO to search for a signal, and found none in the data (see Figs.~\ref{fig:correlation} and \ref{fig:phasefoldedcombined}). Finally, we injected our data with the models to test which models we could recover, thereby placing constraints. The final summary of our results is illustrated in Figs.~\ref{fig:recoverygrid} to \ref{fig:recoverygridfortney}.

Unless the atmosphere is cloudy/hazy, it is evident from the results of our parametric models that the atmosphere of 55 Cnc e cannot have significant presence of water. We rule out lightweight water-rich atmospheres of VMR = 10\% and $\mu$ \textless\ 15 g/mol at a $3\sigma$ confidence level. For cloudless atmospheres with less water content, this lower limit of $\mu$ is relaxed but still quite strong. We can say that even with VMR = 0.1\%, the atmosphere must be heavy with $\mu$ \textgreater\ 10 g/mol. These constraints are stronger than those made in \citetalias{esteves17}, who concluded that for VMR = 10\%, atmospheres with only $\mu$ \textless\ 5 g/mol are ruled out.

Furthermore, we are able to place strong constraints on the presence of TiO using our parametric models. We find that a low mean molecular weight atmosphere would have a VMR of less than $10^{-9}$ with $3\sigma$ confidence. As the atmosphere gets heavier, this constraint is relaxed. For example, an atmosphere with $\mu$ = 10 g/mol would have have a VMR of less than $10^{-7}$ at the $3\sigma$ level.

In contrast to \citetalias{esteves17}, we now use parametric models that match the recent value of planetary radius from \cite{bourrier18} which results in a slightly reduced scale-height and therefore reduced amplitude of the features in the planet's atmosphere.

For a cloudy or hazy atmosphere, the signal would be suppressed even further, and depending on both the pressure level of the cloud tops and the VMR, the features from water could be fully blocked. Therefore our limits are for a cloud-free atmosphere. \cite{mahapatra17} analyze cloud formations on 55 Cnc e and find that mineral clouds may occur, which could explain the featureless results.

Using our fully self-consistent models, our data also revealed no significant signals from an atmosphere with or without the presence of TiO/VO. For the three different solar metallicities used in these models (1x, 10x, 100x), we conclude that atmospheres resembling these models would have been detected at a high significance, indicating that either the planet has a significantly different composition with a much higher mean-molecular weight atmosphere, there is a cloud layer obscuring most of the features, or the planet has no atmosphere at all.

In addition to the species that we have searched for, Ti, Fe, Ti+, and Fe+ also have significant features in the visible spectrum. The first three of these have already been detected using high resolution Doppler spectroscopy in hot Jupiter KELT-9b \citep{hoeijmakers18}, for example. While we have not determined the viability of these species in the the atmosphere of 55 Cnc e, these are potential candidates for future searches in the visible band.

Our results reinforce the findings of \citetalias{esteves17} that the Doppler cross correlation is a very powerful method of recovering signals from nearby super Earths, even though no signal was seen in this particular case. Water signals from such exoplanets are clearly recoverable using ground-based observations. With the launch of the Transiting Exoplanet Survey Satellite (TESS) promising a number nearby follow-up candidates for transit observations, ground based observations may play an increasingly important role in characterizing the atmospheres of super Earths. We expect that with more suitable targets available in the near future, we will have a much more complete understanding of the nature of such worlds.

\acknowledgements
The authors thank Lisa Esteves and Ryan Cloutier for insightful discussions.

This work was based on observations obtained with ESPaDOnS, located at the Canada-France-Hawaii Telescope (CFHT). CFHT is operated by the National Research Council of Canada, the Institut National des Sciences de l'Univers of the Centre National de la Recherche Scientique of France, and the University of Hawai'i. ESPaDOnS is a collaborative project funded by France (CNRS, MENESR, OMP, LATT), Canada (NSERC), CFHT and ESA. ESPaDOnS was remotely controlled from the Gemini Observatory, which is operated by the Association of Universities for Research in Astronomy, Inc., under a cooperative agreement with the NSF on behalf of the Gemini partnership: the National Science Foundation (United States), the National Research Council (Canada), CONICYT (Chile), Ministerio de Ciencia, Tecnología e Innovación Productiva (Argentina) and Ministério da Ciência, Tecnologia e Inovação (Brazil).

\appendix

\begin{figure*}
\includegraphics[width=\textwidth]{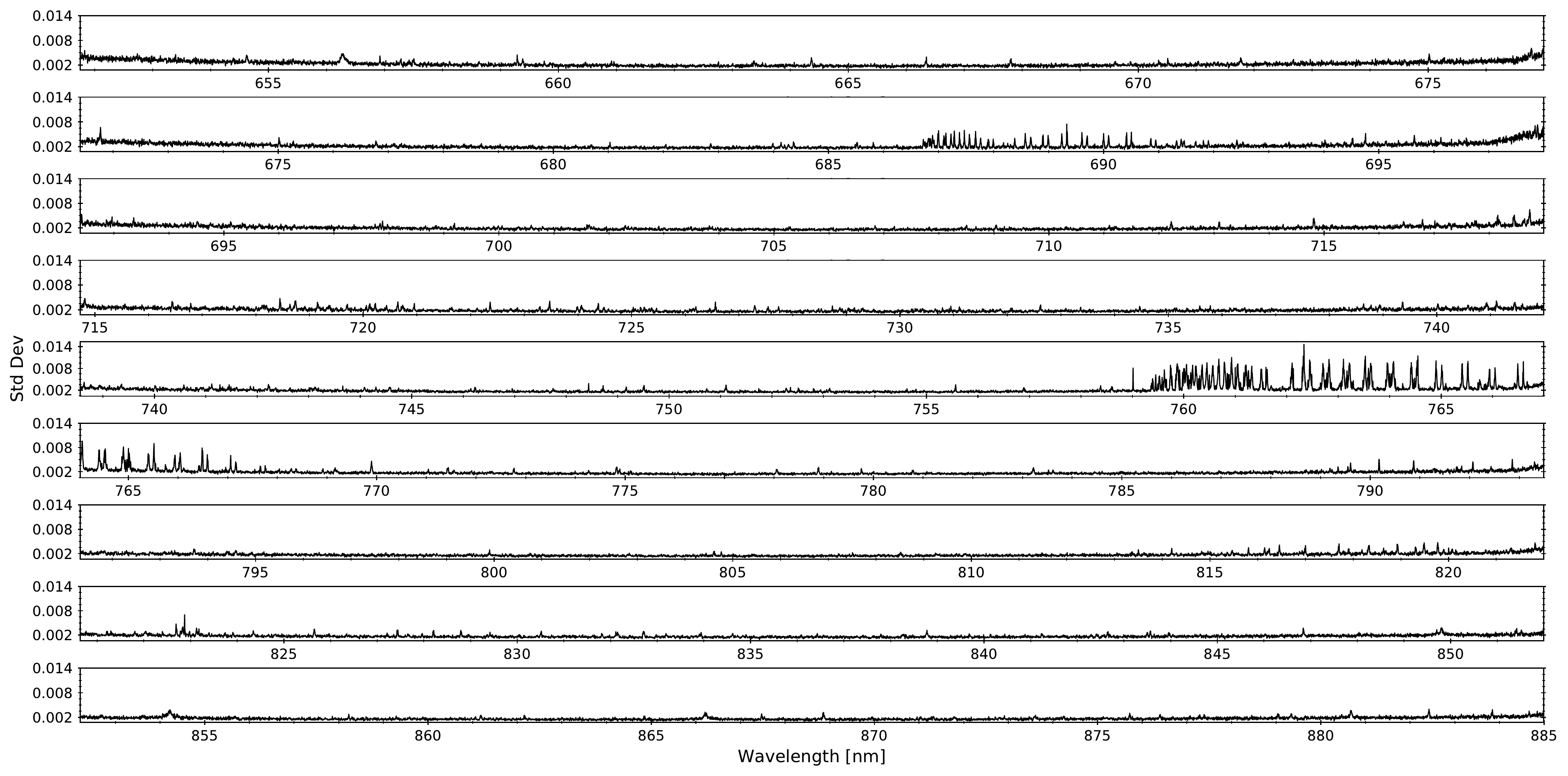}
\caption{Standard deviation of residuals across frames for nine wavelength orders (extended version of bottom panel of Fig.~\ref{fig:ReductionSteps}) for N5.
\label{fig:ExtendedStd1}}
\end{figure*}

\begin{figure*}
\includegraphics[width=\textwidth]{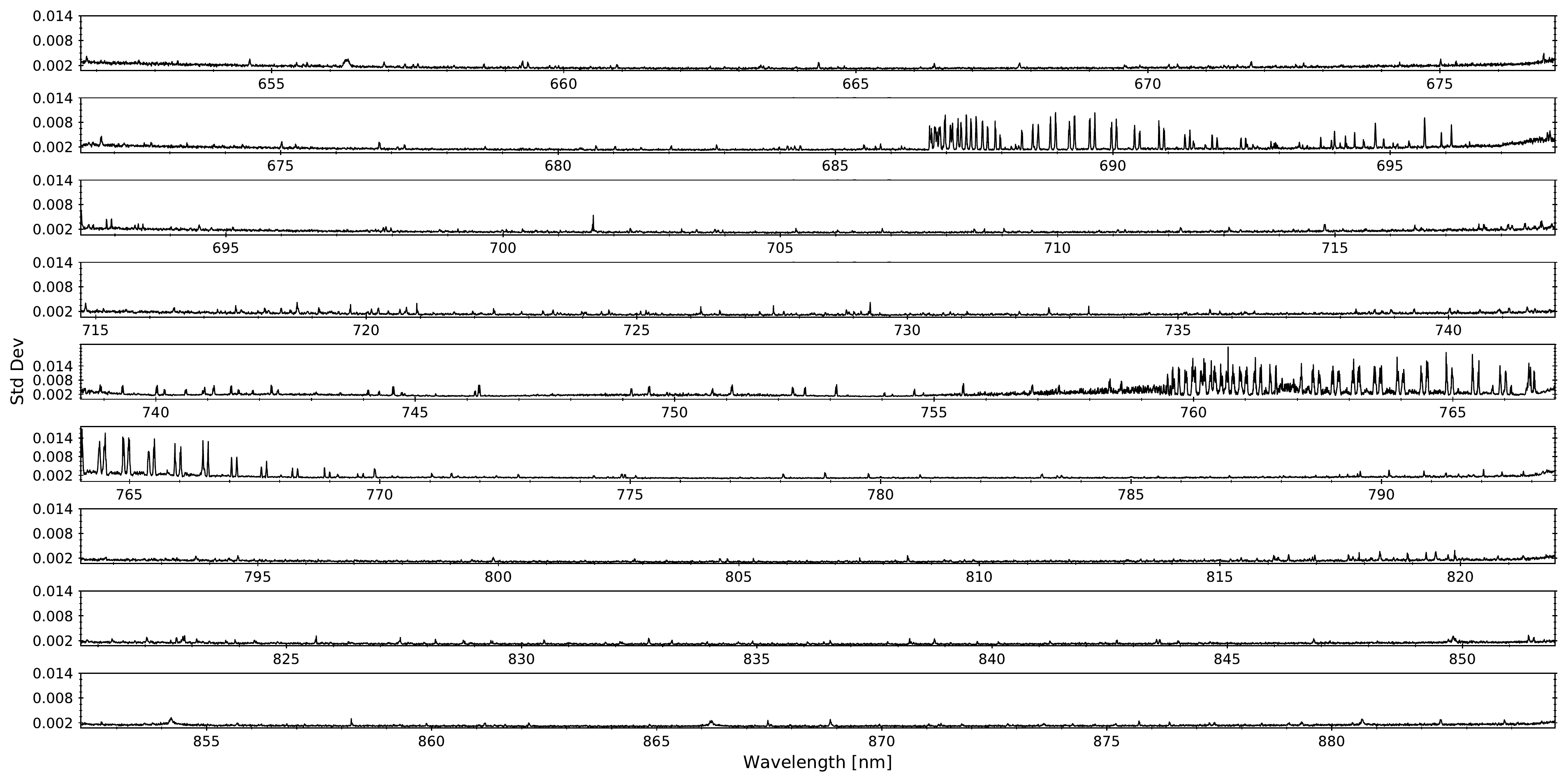}
\caption{Standard deviation of residuals across frames for nine wavelength orders (extended version of bottom panel of Fig.~\ref{fig:ReductionSteps}) for N6.
\label{fig:ExtendedStd2}}
\end{figure*}

\begin{figure*}
\includegraphics[width=\textwidth]{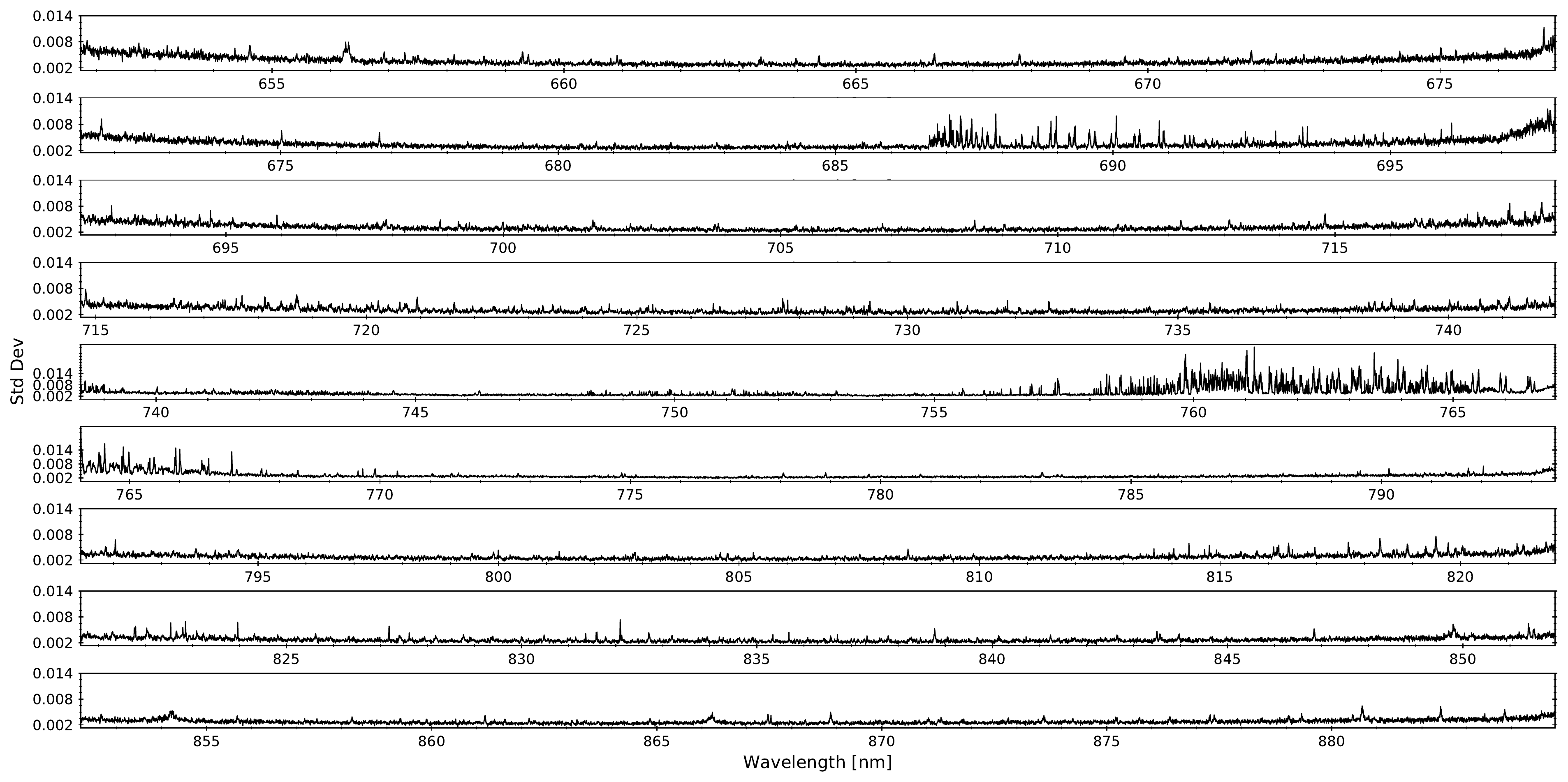}
\caption{Standard deviation of residuals across frames for nine wavelength orders (extended version of bottom panel of Fig.~\ref{fig:ReductionSteps}) for N7.
\label{fig:ExtendedStd3}}
\end{figure*}

\begin{figure*}
\includegraphics[width=\textwidth]{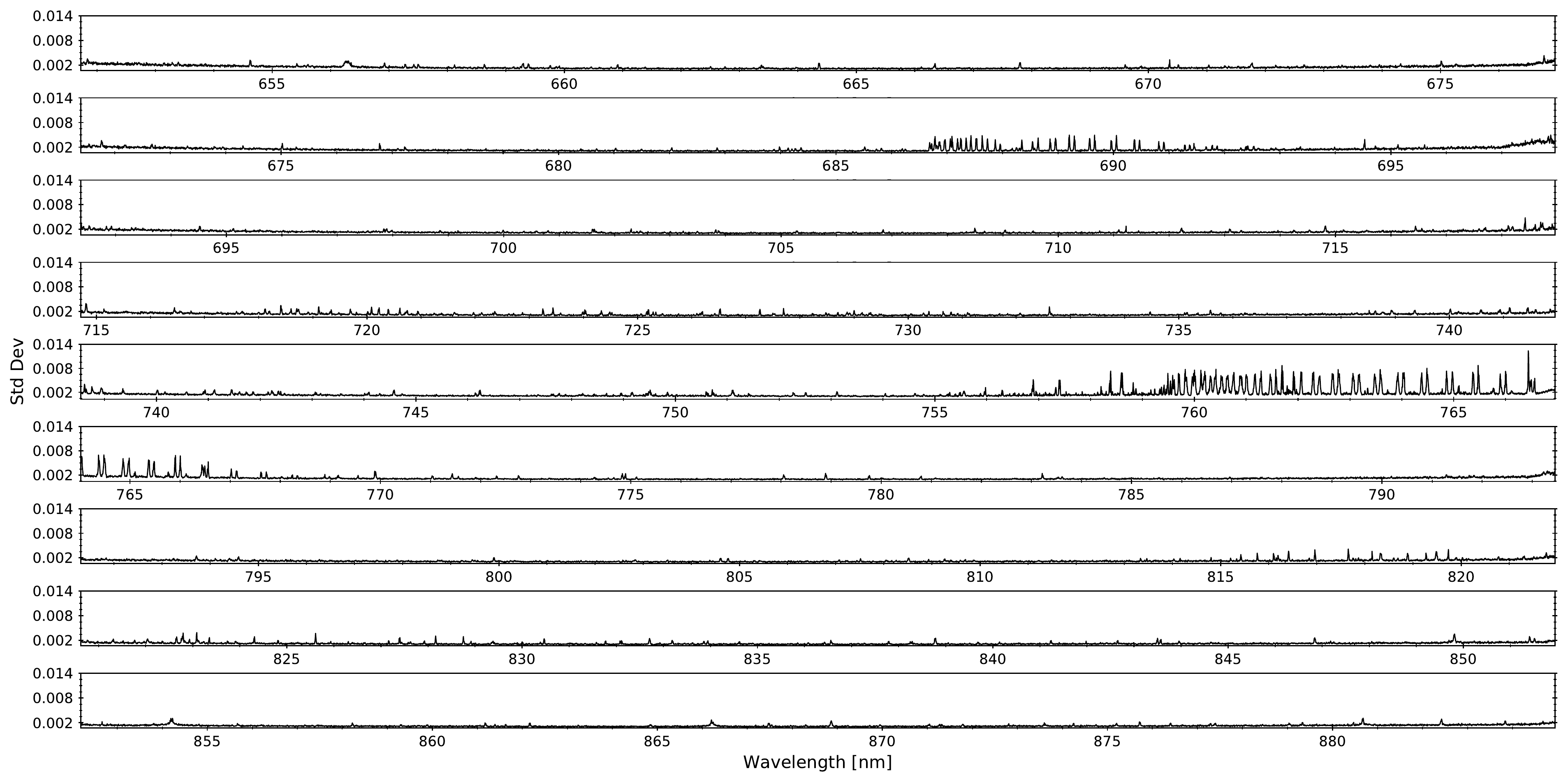}
\caption{Standard deviation of residuals across frames for nine wavelength orders (extended version of bottom panel of Fig.~\ref{fig:ReductionSteps}) for N8.
\label{fig:ExtendedStd4}}
\end{figure*}

\renewcommand\thefigure{\thesection{A}\arabic{figure}}
\setcounter{figure}{0}

\bibliographystyle{aasjournal}
\bibliography{Jindal55Cnce}

\begin{thebibliography}{}
\expandafter\ifx\csname natexlab\endcsname\relax\def\natexlab#1{#1}\fi
\providecommand{\url}[1]{\href{#1}{#1}}
\providecommand{\dodoi}[1]{doi:~\href{http://doi.org/#1}{\nolinkurl{#1}}}
\providecommand{\doeprint}[1]{\href{http://ascl.net/#1}{\nolinkurl{http://ascl.net/#1}}}
\providecommand{\doarXiv}[1]{\href{https://arxiv.org/abs/#1}{\nolinkurl{https://arxiv.org/abs/#1}}}

\bibitem[{Allard {et~al.}(2016)Allard, Spiegelman, \& Kielkopf}]{allard16}
Allard, N., Spiegelman, F., \& Kielkopf, J. 2016, Astronomy \& Astrophysics,
  589, A21

\bibitem[{Angelo \& Hu(2017)}]{angelo17}
Angelo, I., \& Hu, R. 2017, The Astronomical Journal, 154, 232

\bibitem[{Barber {et~al.}(2006)Barber, Tennyson, Harris, \&
  Tolchenov}]{barber06}
Barber, R., Tennyson, J., Harris, G.~J., \& Tolchenov, R. 2006, Monthly Notices
  of the Royal Astronomical Society, 368, 1087

\bibitem[{{Barnes} {et~al.}(2008){Barnes}, {Barman}, {Jones}, {Leigh}, {Collier
  Cameron}, {Barber}, \& {Pinfield}}]{barnes08}
{Barnes}, J.~R., {Barman}, T.~S., {Jones}, H.~R.~A., {et~al.} 2008, \mnras,
  390, 1258, \dodoi{10.1111/j.1365-2966.2008.13831.x}

\bibitem[{{Barnes} {et~al.}(2007{\natexlab{a}}){Barnes}, {Barman}, {Prato},
  {Segransan}, {Jones}, {Leigh}, {Collier Cameron}, \& {Pinfield}}]{barnes07b}
{Barnes}, J.~R., {Barman}, T.~S., {Prato}, L., {et~al.} 2007{\natexlab{a}},
  \mnras, 382, 473, \dodoi{10.1111/j.1365-2966.2007.12394.x}

\bibitem[{{Barnes} {et~al.}(2007{\natexlab{b}}){Barnes}, {Leigh}, {Jones},
  {Barman}, {Pinfield}, {Collier Cameron}, \& {Jenkins}}]{barnes07a}
{Barnes}, J.~R., {Leigh}, C.~J., {Jones}, H.~R.~A., {et~al.}
  2007{\natexlab{b}}, \mnras, 379, 1097,
  \dodoi{10.1111/j.1365-2966.2007.11990.x}

\bibitem[{{Bean} {et~al.}(2010){Bean}, {Miller-Ricci Kempton}, \&
  {Homeier}}]{bean10}
{Bean}, J.~L., {Miller-Ricci Kempton}, E., \& {Homeier}, D. 2010, \nat, 468,
  669, \dodoi{10.1038/nature09596}

\bibitem[{{Bean} {et~al.}(2011){Bean}, {D{\'e}sert}, {Kabath}, {Stalder},
  {Seager}, {Miller-Ricci Kempton}, {Berta}, {Homeier}, {Walsh}, \&
  {Seifahrt}}]{bean11}
{Bean}, J.~L., {D{\'e}sert}, J.-M., {Kabath}, P., {et~al.} 2011, \apj, 743, 92,
  \dodoi{10.1088/0004-637X/743/1/92}

\bibitem[{{Benneke} {et~al.}(2019{\natexlab{a}}){Benneke}, {Knutson},
  {Lothringer}, {Crossfield}, {Moses}, {Morley}, {Kreidberg}, {Fulton},
  {Dragomir}, {Howard}, {Wong}, {D{\'e}sert}, {McCullough}, {Kempton},
  {Fortney}, {Gilliland}, {Deming}, \& {Kammer}}]{Benneke19a}
{Benneke}, B., {Knutson}, H.~A., {Lothringer}, J., {et~al.} 2019{\natexlab{a}},
  Nature Astronomy, 3, 813, \dodoi{10.1038/s41550-019-0800-5}

\bibitem[{{Benneke} {et~al.}(2019{\natexlab{b}}){Benneke}, {Wong}, {Piaulet},
  {Knutson}, {Crossfield}, {Lothringer}, {Morley}, {Gao}, {Greene}, {Dressing},
  {Dragomir}, {Howard}, {McCullough}, {Fortney}, \& {Fraine}}]{Benneke19b}
{Benneke}, B., {Wong}, I., {Piaulet}, C., {et~al.} 2019{\natexlab{b}}, arXiv
  e-prints.
\newblock \doarXiv{1909.04642}

\bibitem[{{Berta} {et~al.}(2012){Berta}, {Charbonneau}, {D{\'e}sert},
  {Miller-Ricci Kempton}, {McCullough}, {Burke}, {Fortney}, {Irwin}, {Nutzman},
  \& {Homeier}}]{berta12}
{Berta}, Z.~K., {Charbonneau}, D., {D{\'e}sert}, J.-M., {et~al.} 2012, \apj,
  747, 35, \dodoi{10.1088/0004-637X/747/1/35}

\bibitem[{{Birkby}(2018)}]{birkby18}
{Birkby}, J.~L. 2018, {Spectroscopic Direct Detection of Exoplanets}, 16

\bibitem[{Borysow(2002)}]{Borysow02}
Borysow, A. 2002, Astronomy \& Astrophysics, 390, 779

\bibitem[{Borysow {et~al.}(2001)Borysow, J{\o}rgensen, \& Fu}]{Borysow01}
Borysow, A., J{\o}rgensen, U.~G., \& Fu, Y. 2001, Journal of Quantitative
  Spectroscopy and Radiative Transfer, 68, 235

\bibitem[{Bourrier {et~al.}(2018)Bourrier, Dumusque, Dorn, Henry,
  Astudillo-Defru, Rey, Benneke, Hebrard, Lovis, Demory, {et~al.}}]{bourrier18}
Bourrier, V., Dumusque, X., Dorn, C., {et~al.} 2018, arXiv preprint
  arXiv:1807.04301

\bibitem[{{Charbonneau} {et~al.}(1998){Charbonneau}, {Jha}, \&
  {Noyes}}]{charbonneau98}
{Charbonneau}, D., {Jha}, S., \& {Noyes}, R.~W. 1998, \apjl, 507, L153,
  \dodoi{10.1086/311703}

\bibitem[{{Charbonneau} {et~al.}(1999){Charbonneau}, {Noyes}, {Korzennik},
  {Nisenson}, {Jha}, {Vogt}, \& {Kibrick}}]{charbonneau99}
{Charbonneau}, D., {Noyes}, R.~W., {Korzennik}, S.~G., {et~al.} 1999, \apjl,
  522, L145, \dodoi{10.1086/312234}

\bibitem[{Chen{\'e} {et~al.}(2014)Chen{\'e}, Pazder, Barrick, Anthony,
  Benedict, Duncan, Gigoux, Kleinman, Malo, Martioli, {et~al.}}]{chene14}
Chen{\'e}, A.-N., Pazder, J., Barrick, G., {et~al.} 2014, in Advances in
  Optical and Mechanical Technologies for Telescopes and Instrumentation, Vol.
  9151, International Society for Optics and Photonics, 915147

\bibitem[{{Croll} {et~al.}(2011){Croll}, {Albert}, {Jayawardhana},
  {Miller-Ricci Kempton}, {Fortney}, {Murray}, \& {Neilson}}]{croll11}
{Croll}, B., {Albert}, L., {Jayawardhana}, R., {et~al.} 2011, \apj, 736, 78,
  \dodoi{10.1088/0004-637X/736/2/78}

\bibitem[{{Crossfield} {et~al.}(2011){Crossfield}, {Barman}, \&
  {Hansen}}]{crossfield11}
{Crossfield}, I.~J.~M., {Barman}, T., \& {Hansen}, B. M.~S. 2011, \apj, 736,
  132, \dodoi{10.1088/0004-637X/736/2/132}

\bibitem[{{Dawson} \& {Fabrycky}(2010)}]{dawson10}
{Dawson}, R.~I., \& {Fabrycky}, D.~C. 2010, \apj, 722, 937,
  \dodoi{10.1088/0004-637X/722/1/937}

\bibitem[{de~Kok {et~al.}(2014)de~Kok, Birkby, Brogi, Schwarz, Albrecht,
  de~Mooij, \& Snellen}]{dekok14}
de~Kok, R.~J., Birkby, J., Brogi, M., {et~al.} 2014, Astronomy \& Astrophysics,
  561, A150

\bibitem[{{de Mooij} {et~al.}(2012){de Mooij}, {Brogi}, {de Kok},
  {Koppenhoefer}, {Nefs}, {Snellen}, {Greiner}, {Hanse}, {Heinsbroek}, {Lee},
  \& {van der Werf}}]{demooij12}
{de Mooij}, E.~J.~W., {Brogi}, M., {de Kok}, R.~J., {et~al.} 2012, \aap, 538,
  A46, \dodoi{10.1051/0004-6361/201117205}

\bibitem[{{Delgado Mena} {et~al.}(2010){Delgado Mena}, {Israelian},
  {Gonz{\'a}lez Hern{\'a}ndez}, {Bond}, {Santos}, {Udry}, \&
  {Mayor}}]{delgadomena10}
{Delgado Mena}, E., {Israelian}, G., {Gonz{\'a}lez Hern{\'a}ndez}, J.~I.,
  {et~al.} 2010, \apj, 725, 2349, \dodoi{10.1088/0004-637X/725/2/2349}

\bibitem[{Demory {et~al.}(2011)Demory, Gillon, Deming, Valencia, Seager,
  Benneke, Lovis, Cubillos, Harrington, Stevenson, {et~al.}}]{demory11}
Demory, B.-O., Gillon, M., Deming, D., {et~al.} 2011, Astronomy \&
  Astrophysics, 533, A114

\bibitem[{Demory {et~al.}(2016)Demory, Gillon, de~Wit, Madhusudhan, Bolmont,
  Heng, Kataria, Lewis, Hu, Krick, {et~al.}}]{demory16}
Demory, B.-O., Gillon, M., de~Wit, J., {et~al.} 2016, Nature, 532, 207

\bibitem[{{Diamond-Lowe} {et~al.}(2018){Diamond-Lowe}, {Berta-Thompson},
  {Charbonneau}, \& {Kempton}}]{diamond18}
{Diamond-Lowe}, H., {Berta-Thompson}, Z., {Charbonneau}, D., \& {Kempton}, E.
  M.~R. 2018, \aj, 156, 42, \dodoi{10.3847/1538-3881/aac6dd}

\bibitem[{{Dorn} {et~al.}(2019){Dorn}, {Harrison}, {Bonsor}, \&
  {Hands}}]{dorn19}
{Dorn}, C., {Harrison}, J.~H.~D., {Bonsor}, A., \& {Hands}, T.~O. 2019, \mnras,
  484, 712, \dodoi{10.1093/mnras/sty3435}

\bibitem[{Esteves {et~al.}(2017)Esteves, De~Mooij, Jayawardhana, Watson, \&
  de~Kok}]{esteves17}
Esteves, L.~J., De~Mooij, E.~J., Jayawardhana, R., Watson, C., \& de~Kok, R.
  2017, The Astronomical Journal, 153, 11pp

\bibitem[{Fortney {et~al.}(2008)Fortney, Lodders, Marley, \&
  Freedman}]{fortney08}
Fortney, J.~J., Lodders, K., Marley, M.~S., \& Freedman, R.~S. 2008, The
  Astrophysical Journal, 678, 1419

\bibitem[{Fortney {et~al.}(2005)Fortney, Marley, Lodders, Saumon, \&
  Freedman}]{fortney05}
Fortney, J.~J., Marley, M., Lodders, K., Saumon, D., \& Freedman, R. 2005, The
  Astrophysical Journal Letters, 627, L69

\bibitem[{Fressin {et~al.}(2013)Fressin, Torres, Charbonneau, Bryson,
  Christiansen, Dressing, Jenkins, Walkowicz, \& Batalha}]{fressin13}
Fressin, F., Torres, G., Charbonneau, D., {et~al.} 2013, The Astrophysical
  Journal, 766, 81

\bibitem[{Fulton {et~al.}(2017)Fulton, Petigura, Howard, Isaacson, Marcy,
  Cargile, Hebb, Weiss, Johnson, Morton, {et~al.}}]{fulton17}
Fulton, B.~J., Petigura, E.~A., Howard, A.~W., {et~al.} 2017, The Astronomical
  Journal, 154, 109

\bibitem[{{Gelman} {et~al.}(2011){Gelman}, {Elkins-Tanton}, \&
  {Seager}}]{gelman11}
{Gelman}, S.~E., {Elkins-Tanton}, L.~T., \& {Seager}, S. 2011, \apj, 735, 72,
  \dodoi{10.1088/0004-637X/735/2/72}

\bibitem[{Gillon {et~al.}(2012)Gillon, Demory, Benneke, Valencia, Deming,
  Seager, Lovis, Mayor, Pepe, Queloz, {et~al.}}]{gillon12}
Gillon, M., Demory, B.-O., Benneke, B., {et~al.} 2012, Astronomy \&
  Astrophysics, 539, A28

\bibitem[{{Hammond} \& {Pierrehumbert}(2017)}]{hammond17}
{Hammond}, M., \& {Pierrehumbert}, R.~T. 2017, \apj, 849, 152,
  \dodoi{10.3847/1538-4357/aa9328}

\bibitem[{{Henning} {et~al.}(2009){Henning}, {O'Connell}, \&
  {Sasselov}}]{henning09}
{Henning}, W.~G., {O'Connell}, R.~J., \& {Sasselov}, D.~D. 2009, \apj, 707,
  1000, \dodoi{10.1088/0004-637X/707/2/1000}

\bibitem[{Hoeijmakers {et~al.}(2015)Hoeijmakers, de~Kok, Snellen, Brogi,
  Birkby, \& Schwarz}]{hoeijmakers15}
Hoeijmakers, H., de~Kok, R., Snellen, I., {et~al.} 2015, Astronomy \&
  Astrophysics, 575, A20

\bibitem[{Hoeijmakers {et~al.}(2018)Hoeijmakers, Ehrenreich, Heng, Kitzmann,
  Grimm, Allart, Deitrick, Wyttenbach, Oreshenko, Pino,
  {et~al.}}]{hoeijmakers18}
Hoeijmakers, H.~J., Ehrenreich, D., Heng, K., {et~al.} 2018, Nature, 560, 453

\bibitem[{{Kite} {et~al.}(2016){Kite}, {Fegley}, {Schaefer}, \&
  {Gaidos}}]{kite16}
{Kite}, E.~S., {Fegley}, Bruce, J., {Schaefer}, L., \& {Gaidos}, E. 2016, \apj,
  828, 80, \dodoi{10.3847/0004-637X/828/2/80}

\bibitem[{Knutson {et~al.}(2014)Knutson, Benneke, Deming, \&
  Homeier}]{knutson14a}
Knutson, H.~A., Benneke, B., Deming, D., \& Homeier, D. 2014, Nature, 505, 66

\bibitem[{Kreidberg {et~al.}(2014)Kreidberg, Bean, D{\'e}sert, Benneke, Deming,
  Stevenson, Seager, Berta-Thompson, Seifahrt, \& Homeier}]{kreidberg14}
Kreidberg, L., Bean, J.~L., D{\'e}sert, J.-M., {et~al.} 2014, Nature, 505, 69

\bibitem[{{Kuchner}(2003)}]{kuchner03}
{Kuchner}, M.~J. 2003, \apjl, 596, L105, \dodoi{10.1086/378397}

\bibitem[{Lammer {et~al.}(2013)Lammer, Erkaev, Odert, Kislyakova, Leitzinger,
  \& Khodachenko}]{lammer13}
Lammer, H., Erkaev, N., Odert, P., {et~al.} 2013, Monthly Notices of the Royal
  Astronomical Society, 430, 1247

\bibitem[{{L{\'e}ger} {et~al.}(2004){L{\'e}ger}, {Selsis}, {Sotin}, {Guillot},
  {Despois}, {Mawet}, {Ollivier}, {Lab{\`e}que}, {Valette}, {Brachet},
  {Chazelas}, \& {Lammer}}]{leger04}
{L{\'e}ger}, A., {Selsis}, F., {Sotin}, C., {et~al.} 2004, \icarus, 169, 499,
  \dodoi{10.1016/j.icarus.2004.01.001}

\bibitem[{{Madhusudhan} {et~al.}(2012){Madhusudhan}, {Lee}, \&
  {Mousis}}]{madhusudhan12}
{Madhusudhan}, N., {Lee}, K. K.~M., \& {Mousis}, O. 2012, \apjl, 759, L40,
  \dodoi{10.1088/2041-8205/759/2/L40}

\bibitem[{Mahapatra {et~al.}(2017)Mahapatra, Helling, \& Miguel}]{mahapatra17}
Mahapatra, G., Helling, C., \& Miguel, Y. 2017, Monthly Notices of the Royal
  Astronomical Society, 472, 447

\bibitem[{Martioli {et~al.}(2012)Martioli, Teeple, Manset, Devost, Withington,
  Venne, \& Tannock}]{martioli12}
Martioli, E., Teeple, D., Manset, N., {et~al.} 2012, in Software and
  Cyberinfrastructure for Astronomy II, Vol. 8451, International Society for
  Optics and Photonics, 84512B

\bibitem[{McArthur {et~al.}(2004)McArthur, Endl, Cochran,
  {et~al.}}]{mcarthur04}
McArthur, B., Endl, M., Cochran, W., {et~al.} 2004, Astrophys. J, 614, L81

\bibitem[{McKemmish {et~al.}(2016)McKemmish, Yurchenko, \&
  Tennyson}]{mckemmish16}
McKemmish, L.~K., Yurchenko, S.~N., \& Tennyson, J. 2016, Monthly Notices of
  the Royal Astronomical Society, 463, 771

\bibitem[{{Modirrousta-Galian} {et~al.}(2020){Modirrousta-Galian}, {Locci},
  {Tinetti}, \& {Micela}}]{modirrousta-galian20}
{Modirrousta-Galian}, D., {Locci}, D., {Tinetti}, G., \& {Micela}, G. 2020,
  \apj, 888, 87, \dodoi{10.3847/1538-4357/ab616b}

\bibitem[{Morley {et~al.}(2017)Morley, Kreidberg, Rustamkulov, Robinson, \&
  Fortney}]{morley17}
Morley, C.~V., Kreidberg, L., Rustamkulov, Z., Robinson, T., \& Fortney, J.~J.
  2017, The Astrophysical Journal, 850, 121

\bibitem[{Nidever {et~al.}(2002)Nidever, Marcy, Butler, Fischer, \&
  Vogt}]{nidever02}
Nidever, D.~L., Marcy, G.~W., Butler, R.~P., Fischer, D.~A., \& Vogt, S.~S.
  2002, The Astrophysical Journal Supplement Series, 141, 503

\bibitem[{Nugroho {et~al.}(2017)Nugroho, Kawahara, Masuda, Hirano, Kotani, \&
  Tajitsu}]{nugroho17}
Nugroho, S.~K., Kawahara, H., Masuda, K., {et~al.} 2017, The Astronomical
  Journal, 154, 221

\bibitem[{Plez(1998)}]{plez98}
Plez, B. 1998, Astronomy and Astrophysics, 337, 495

\bibitem[{Ridden-Harper {et~al.}(2016)Ridden-Harper, Snellen, Keller, De~Kok,
  Di~Gloria, Hoeijmakers, Brogi, Fridlund, Vermeersen, \& van
  Westrenen}]{ridden-harper16}
Ridden-Harper, A., Snellen, I., Keller, C., {et~al.} 2016, Astronomy \&
  Astrophysics, 593, A129

\bibitem[{{Rodler} {et~al.}(2008){Rodler}, {K{\"u}rster}, \&
  {Henning}}]{rodler08}
{Rodler}, F., {K{\"u}rster}, M., \& {Henning}, T. 2008, \aap, 485, 859,
  \dodoi{10.1051/0004-6361:20079175}

\bibitem[{{Rogers} \& {Seager}(2010)}]{rogers10}
{Rogers}, L.~A., \& {Seager}, S. 2010, \apj, 716, 1208,
  \dodoi{10.1088/0004-637X/716/2/1208}

\bibitem[{Rothman {et~al.}(2010)Rothman, Gordon, Barber, Dothe, Gamache,
  Goldman, Perevalov, Tashkun, \& Tennyson}]{rothman10}
Rothman, L., Gordon, I., Barber, R., {et~al.} 2010, Journal of Quantitative
  Spectroscopy and Radiative Transfer, 111, 2139

\bibitem[{{Schaefer} \& {Fegley}(2009)}]{schaefer09}
{Schaefer}, L., \& {Fegley}, B. 2009, \apjl, 703, L113,
  \dodoi{10.1088/0004-637X/703/2/L113}

\bibitem[{Schwenke(1998)}]{schwenke98}
Schwenke, D.~W. 1998, Faraday Discussions, 109, 321

\bibitem[{Snellen {et~al.}(2010)Snellen, De~Kok, De~Mooij, \&
  Albrecht}]{snellen10}
Snellen, I.~A., De~Kok, R.~J., De~Mooij, E.~J., \& Albrecht, S. 2010, Nature,
  465, 1049

\bibitem[{{Sotin} {et~al.}(2007){Sotin}, {Grasset}, \& {Mocquet}}]{sotin07}
{Sotin}, C., {Grasset}, O., \& {Mocquet}, A. 2007, \icarus, 191, 337,
  \dodoi{10.1016/j.icarus.2007.04.006}

\bibitem[{{Southworth} {et~al.}(2017){Southworth}, {Mancini}, {Madhusudhan},
  {Molli{\`e}re}, {Ciceri}, \& {Henning}}]{southworth17}
{Southworth}, J., {Mancini}, L., {Madhusudhan}, N., {et~al.} 2017, \aj, 153,
  191, \dodoi{10.3847/1538-3881/aa6477}

\bibitem[{Tamuz {et~al.}(2005)Tamuz, Mazeh, \& Zucker}]{tamuz05}
Tamuz, O., Mazeh, T., \& Zucker, S. 2005, Monthly Notices of the Royal
  Astronomical Society, 356, 1466

\bibitem[{Teeple(2014)}]{teeple14}
Teeple, D. 2014, Astrophysics Source Code Library

\bibitem[{Tennyson \& Yurchenko(2012)}]{tennyson12}
Tennyson, J., \& Yurchenko, S.~N. 2012, Monthly Notices of the Royal
  Astronomical Society, 425, 21

\bibitem[{{Teske} {et~al.}(2013){Teske}, {Cunha}, {Schuler}, {Griffith}, \&
  {Smith}}]{teske13}
{Teske}, J.~K., {Cunha}, K., {Schuler}, S.~C., {Griffith}, C.~A., \& {Smith},
  V.~V. 2013, \apj, 778, 132, \dodoi{10.1088/0004-637X/778/2/132}

\bibitem[{{Tsiaras} {et~al.}(2019){Tsiaras}, {Waldmann}, {Tinetti}, {Tennyson},
  \& {Yurchenko}}]{Tsiaras19}
{Tsiaras}, A., {Waldmann}, I.~P., {Tinetti}, G., {Tennyson}, J., \&
  {Yurchenko}, S.~N. 2019, Nature Astronomy, \dodoi{10.1038/s41550-019-0878-9}

\bibitem[{Tsiaras {et~al.}(2016)Tsiaras, Rocchetto, Waldmann, Venot, Varley,
  Morello, Damiano, Tinetti, Barton, Yurchenko, {et~al.}}]{tsiaras16}
Tsiaras, A., Rocchetto, M., Waldmann, I., {et~al.} 2016, The Astrophysical
  Journal, 820, 99

\bibitem[{{Tsiaras} {et~al.}(2018){Tsiaras}, {Waldmann}, {Zingales},
  {Rocchetto}, {Morello}, {Damiano}, {Karpouzas}, {Tinetti}, {McKemmish},
  {Tennyson}, \& {Yurchenko}}]{Tsiaras18}
{Tsiaras}, A., {Waldmann}, I.~P., {Zingales}, T., {et~al.} 2018, \aj, 155, 156,
  \dodoi{10.3847/1538-3881/aaaf75}

\bibitem[{Valencia {et~al.}(2007)Valencia, Sasselov, \& O'Connell}]{valencia07}
Valencia, D., Sasselov, D.~D., \& O'Connell, R.~J. 2007, The Astrophysical
  Journal, 656, 545

\bibitem[{{Wiedemann}(1996)}]{wiedemann96}
{Wiedemann}, G. 1996, The Messenger, 86, 24

\bibitem[{{Wiedemann} {et~al.}(2001){Wiedemann}, {Deming}, \&
  {Bjoraker}}]{wiedemann01}
{Wiedemann}, G., {Deming}, D., \& {Bjoraker}, G. 2001, \apj, 546, 1068,
  \dodoi{10.1086/318316}

\bibitem[{Winn {et~al.}(2011)Winn, Matthews, Dawson, Fabrycky, Holman,
  Kallinger, Kuschnig, Sasselov, Dragomir, Guenther, {et~al.}}]{winn11}
Winn, J.~N., Matthews, J.~M., Dawson, R.~I., {et~al.} 2011, The Astrophysical
  Journal Letters, 737, L18

\end{thebibliography}

%%%%%%%%%%%%%%%%%%%%%%%%%%%%%%%%%%%%%%%%%%%%%%%%%%%

\end{document}